\documentclass[letterpaper, 10 pt, conference]{ieeeconf}

\IEEEoverridecommandlockouts
\usepackage{amsmath}
\usepackage{amssymb}
\usepackage{tabularx}
\usepackage{booktabs}
\usepackage{bm}
\usepackage{color}
\usepackage{units}
\usepackage{graphics}
\usepackage{graphicx}
\usepackage{float}
\usepackage{xcolor}

\usepackage{tikz}
\usetikzlibrary{shapes.geometric,arrows}
\usetikzlibrary{calc}
\usetikzlibrary{positioning}
\usepackage{pgfplots}
\pgfplotsset{
	legend style = {font = {\fontsize{1.8ex}{12 pt}\selectfont}, 
	fill = none,
	draw=none
	},  
}
\pgfplotsset{compat=1.16} 

\usepackage[labelformat=simple]{subcaption}  

\newenvironment{minimatrix}{\left[\begin{large}\begin{smallmatrix}}{\end{smallmatrix}\end{large}\right]}

\usepackage{hyperref}
\usepackage{cleveref}
\crefname{figure}{Fig.}{Figs.}
\crefname{table}{Table}{Tables}
\crefname{equation}{Eq.}{Eqs.}
\Crefname{equation}{Equation}{Equations}
\crefname{chapter}{Chapter}{Chapters}
\crefname{appendix}{Appendix}{Appendices}
\crefname{section}{Section}{Sections}
\crefname{subsection}{Section}{Sections}
\crefname{subsubsection}{Section}{Sections}

\title{\LARGE \bf
Data-driven Nonlinear Model Reduction using Koopman Theory: Integrated Control Form and NMPC Case Study
}

\author{Jan C.~Schulze and Alexander Mitsos%
\thanks{J.C.S.~and A.M.~are with the Chair of Process Systems Engineering (AVT.SVT), 
Department of Mechanical Engineering,
RWTH Aachen University, Germany.
Correspondence:
{\tt\footnotesize 
amitsos@alum.mit.edu}}
}

\newcommand\copyrighttext{\footnotesize 
Accepted manuscript. Final publication in IEEE Control Systems Letters, Vol.~6, 2022, DOI: 10.1109/LCSYS.2022.3181443\\
© 2022 IEEE. Personal use of this material is permitted. Permission from IEEE must be obtained for all other uses, in any current or future
media, including reprinting/republishing this material for advertising or promotional purposes, creating new collective works, for resale or
redistribution to servers or lists, or reuse of any copyrighted component of this work in other works.
  }
\newcommand\copyrightnotice{%
\begin{tikzpicture}[remember picture,overlay]
\node[anchor=south,yshift=10pt] at (current page.south) {\fbox{\parbox{\dimexpr\textwidth-\fboxsep-\fboxrule\relax}{\copyrighttext}}};
\end{tikzpicture}%
}

\begin{document}

\maketitle
\thispagestyle{empty}
\pagestyle{empty}

\begin{abstract}
We use Koopman theory for data-driven model reduction of nonlinear dynamical systems with controls.
We propose generic model structures combining delay-coordinate encoding of measurements and full-state decoding to integrate reduced Koopman modeling and state estimation.
We present a deep-learning approach to train the proposed models.
A case study demonstrates that our approach provides accurate control models and enables real-time capable nonlinear model predictive control of a high-purity cryogenic distillation column.
\end{abstract}


\section{Introduction}\copyrightnotice
\label{sec:intro}
Model order reduction is a powerful technique to achieve real-time nonlinear model predictive control (NMPC) when using large-scale process models \cite{Marquardt.2002}.
Reduction methods construct low-order approximations of high-order differential equations by projecting the dynamical system from the full state space to a lower dimensional subspace \cite{Antoulas.2005,Benner.2017}. 
While reduction methods for linear dynamics are relatively mature, 
nonlinear model reduction is an ongoing topic of research, e.g., \cite{Lee.2020}.
Besides general-purpose reduction methods, 
tailored reduced models for specific systems have been derived either from system-theoretic reduction methods, or based on physical insight with simplifying assumptions, e.g., \cite{Benallou.1986, CaspariWave}.
These reduced models often have an intuitive and interpretable structure. 
However, their suitability depends on the individual 
system properties, requiring expert knowledge and additional modeling efforts.

Data-driven reduction approaches are non-intrusive, i.e., build the reduced dynamics directly from simulation data of the high-order model.
Their advantage is the decoupling of the reduction process from the original model and little required expert knowledge of the individual target system.
Among a variety of approaches are the Loewner framework \cite{Antoulas.2017}, data-driven moment matching \cite{Scarciotti.2017}, as well as
Koopman theory \cite{Mezic.2005, Mauroy.2016} and the related (extended) dynamic mode decomposition \cite{Williams.2015}.
Koopman approaches have a system-theoretical foundation and
offer a practicable compromise between simple structure, universal form and high accuracy.
For example, Koopman deep learning combines artificial neural networks (ANNs) and linear dynamics to learn nonlinear dynamics \cite{Lusch.2018}.

While Koopman theory was originally introduced for autonomous systems, extensions to control systems have been proposed  \cite{Surana.2016, Korda.2018}, providing linear or bilinear models.
We  recently presented a Koopman model formulation that realizes a multiple-input multiple-output Wiener structure \cite{Schulze.2022}, and demonstrated its suitability for data-driven model reduction using deep learning.
We summarize these modeling approaches in Section \ref{sec:review}.
A further discussion of learning-based control, such as online learning for continuous model improvement, can be found in the recent review \cite{Hewing.2020}.

Although model reduction lowers the computational cost of optimization, knowledge of the initial states remains a prerequisite for using the model.
However, in practice the full states are typically not available online for process feedback, requiring additional state estimation.
Instead, we are interested in models that can directly cope with missing state information. 
Motivated by Takens' embedding theorem, several works have successfully used time-delay coordinates, i.e., Hankel matrices, to robustify Koopman approaches in cases of missing state information, chaotic or non-Markovian behavior \cite{Svenkeson.2016, Brunton.2017, Das.2019}.
In addition, the delay-embedding approach was adopted for Koopman MPC in \cite{Korda.2018}. 
Related data-driven strategies for linear systems have been developed recently in the context of Willems' fundamental lemma \cite{Willems.2005, Coulson.2019, Schmitz.2022}.
Data-driven model identification based on delayed measurements eliminates the need for state estimation due to the autoregressive character.
However, the delay-embedding models above 
predict future measurements rather than reconstructing the underlying unmeasured states.
Hence, these approaches do not allow control or monitoring of states that cannot be (frequently) measured.

To overcome this issue, we propose a reduced Koopman structure that integrates delay embedding of measurements for initialization and full state reconstruction for prediction.
Thereby, we incorporate state estimation into the data-driven model reduction process by 
``encoding'' a time series of measurements, and ``decoding'' the full state vector. 
We present extensions of both the Linear and Wiener-type Koopman forms to combine these aspects (Section \ref{sec:methods}).
Importantly, the proposed structures directly initialize the latent Koopman coordinates, instead of estimating the original states from measurements.
Our approach differs from other works as it facilitates monitoring and control of unmeasured states (in contrast to, e.g., \cite{Korda.2018}), while applicable without an external state estimator (in contrast to \cite{Lusch.2018, Surana.2016, Schulze.2022}).

We extend our deep-learning model reduction framework \cite{Schulze.2022} to train the proposed models (Section \ref{sec:deeplearning}),
and apply our strategy in a case study for model predictive control of a high-purity cryogenic distillation column (Section \ref{sec:casestudy}).
Therein, we demonstrate the effectiveness of both the reduction approach and the state estimation features
on a high-order example of industrial relevance.

\section{Koopman theory for control}
\label{sec:review}
Koopman theory enables the global linearization of nonlinear autonomous dynamical systems by means of nonlinear coordinate lifting \cite{Mezic.2005, Mauroy.2016}.
The family of Koopman operators acts on a function space of nonlinear observables, which effectively results in a linear but infinite dimensional representation of the original dynamics.
However, similar to truncating a singular value decomposition,
dominant Koopman eigenfunctions can be sought to extract dynamical patterns, 
which facilitates low-order modeling of moderately nonlinear dynamical systems
\cite{Williams.2015, Lusch.2018}. 
Modifications that permit related truncations for stronger nonlinearities have been discussed recently, e.g., \cite{Svenkeson.2016, Brunton.2017}. 

To apply Koopman theory to nonlinear systems with controls, we consider the class of asymptotically stable input-affine systems:
\begin{equation}
	\label{eqn:inputaffine}
	\dot{\bm x}(t) = \bm f (\bm x(t)) + \sum_{i=1}^{n_u} \bm g_i(\bm x(t)) u_i(t) \,,
\end{equation}
where $\bm{x}(t) \in \mathbb{R}^{n_x}$  
are the differential states, 
$\bm{u}(t) \in \mathbb{R}^{n_u}$ 
are external inputs, 
and 
$\bm f: \mathbb{R}^{n_x} \rightarrow \mathbb{R}^{n_x}$, 
$\bm g_i: \mathbb{R}^{n_x} \rightarrow \mathbb{R}^{n_x}$ 
are continuously differentiable vector fields.
This type of non-autonomous systems can be treated by the Koopman operator framework, yielding a global bilinearization \cite{Surana.2016}:
\begin{subequations}
	\label{eqn:bilinear}
	\begin{flalign}
		\dot{\bm z}(t) &= \bar A \bm z(t) + \sum_{i=1}^{n_u} \bar B_i \bm z(t) u_i(t)  \,,\\
		\bm x (t) &= \bar C^x \bm z(t)\,,\\
		\bm z(t_0) &= \bar{\bm \Psi}(\bm x(t_0)) \,.
	\end{flalign}
\end{subequations}
Therein, 
$\bm z(t) \in \mathbb{R}^{n_z}$ 
are the Koopman canonical coordinates \cite{Surana.2016},
$\bar{\bm \Psi}: \mathbb{R}^{n_x} \rightarrow \mathbb{R}^{n_z}$  
represents the nonlinear transformation (lifting) to the Koopman coordinates and also provides the initial conditions, and $\bar A$, $\bar B_i$, $\bar C^x$ are matrices and assumed as constant.
In some cases, the bilinear Koopman dynamics may be simplified to linear dynamics; we refer to Refs.~\cite{Surana.2016, Korda.2018} for further details.

We have recently shown \cite{Schulze.2022} that, alternatively, a Wiener representation may be derived for \cref{eqn:inputaffine}: 
\begin{subequations}
	\label{eqn:wiener}
	\begin{flalign}
		\dot{\bm z}(t) &= A \bm z(t) + B \bm u(t)  \,, \label{eqn:lineardynamics}\\
		\bm x (t) &= \bm T(\bm z(t))\,, \label{eqn:transformation} \\
		\bm z(t_0) &= \bm \Psi(\bm x(t_0)) \,, \label{eqn:initialcondition}
	\end{flalign}
\end{subequations}
where $\bm T: \mathbb{R}^{n_z} \rightarrow \mathbb{R}^{n_x}$ 
describes a continuously invertible nonlinear transformation from the 
Koopman coordinates to the original state space.
Here, we will use \cref{eqn:wiener} as a reduced approximation, 
specifically $n_z < n_x$, 
where $\bm T(\,\cdot\,)$
characterizes the low-dimensional solution manifold of the reduced model from the extrinsic view in state space.
Notice that the Wiener type unifies both model order reduction and simplification (linearization) of the latent dynamics.

In our work \cite{Schulze.2022}, we investigated the Wiener form as a generic model type for data-driven dynamical approximation and reduction.
To this end, we presented a universal deep-learning strategy that generates the matrices and transformations of Eqs.~(\ref{eqn:lineardynamics}) to (\ref{eqn:initialcondition}) simultaneously. 
Throughout different nonlinear examples, we observed a high accuracy of Wiener-type Koopman models.
Importantly, the Wiener form strictly divides the model into linear dynamics and nonlinear transformation, which we regard as favorable in terms of dynamic analysis (e.g.,~stability),
transformation between discrete and continuous time, and potential structural exploitation.
Overall, we consider deep learning of Wiener-type Koopman models as an effective means for non-intrusive data-driven reduction of moderately nonlinear dynamics.

\section{Extensions of the Wiener-type model}
\label{sec:methods}
Herein, we present two extensions of the Wiener-type Koopman form that enable the incorporation of nonlinear system outputs (measurements) and state estimation features.
\subsection{Observer form of input-affine systems}
In many situations, we are interested in modeling nonlinear measurements of the dynamic states as well.
Consider an observer form of \cref{eqn:inputaffine} with nonlinear output relation:
\begin{subequations}
	\label{eqn:observer}
	\begin{flalign}
		\dot{\bm x}(t) &= \bm f (\bm x(t)) + \sum_{i=1}^{n_u} \bm g_i(\bm x(t)) u_i(t) \,,\\
		\bm y(t) &= \bm h(\bm x(t)) \,, \label{eqn:outputs}
	\end{flalign}
\end{subequations}
where $\bm y(t) \in \mathbb{R}^{n_y}$ are the outputs, and $\bm h: \mathbb{R}^{n_x} \rightarrow \mathbb{R}^{n_y}$.
For bilinear Koopman models, Surana \cite{Surana.2016} suggests to select Koopman 
coordinates $\bm z$ such that the outputs $\bm y$ are covered by the expansion \cite{Surana.2016}:
\begin{equation}
	\label{eqn:koopmanmode}
	\bm y(t) = \sum_{k=1}^{n_z} \bm v_k^{(\bm y)} z_k(t) = \bar C^y \bm z (t) \,,
\end{equation}
where $\bm v_k^{(\bm y)} \in \mathbb{R}^{n_y}$ are Koopman canonical modes grouped into a matrix $\bar C^y$.
Here, we draw on a related but less restrictive argument to extend
the Wiener-type Koopman model (\ref{eqn:wiener}).
Due to the nonlinear transformation (\ref{eqn:transformation}), 
we can augment the model: 
\begin{subequations}
	\label{eqn:Koopman_observer}
	\begin{flalign}
		\dot{\bm z}(t) &= A \bm z(t) + B \bm u(t)  \,, \\
		\left[\begin{large}\begin{smallmatrix}\bm x(t) \\ \bm y(t)
		\end{smallmatrix}\end{large}\right]
		&= \hat{\bm T}(\bm z(t))\,,
		\\
		\bm z(t_0) &= \bm \Psi(\bm x(t_0)) \,.
	\end{flalign}
\end{subequations}
Clearly, the straightforward way is to combine $\bm T$ and $\bm h \circ \bm T$ into $\hat{\bm T}$, which does not affect the choice of Koopman coordinates. 
However, with regards to deep-learning approximations, there may exist direct and more ``efficient'' mappings from $\bm z$ to $\bm y$.

\subsection{Delay-embedding for state estimation}
To equip the Wiener-type Koopman model with estimation features, we modify the structure, 
\cref{eqn:Koopman_observer}.
First, we construct delay-coordinates $\bm \chi$ as a series of time-shifted equidistant measurements that we arrange in a Hankel matrix:
\begin{equation}
	\mathcal{H} =
	\left[
	\begin{large}
		\begin{smallmatrix}
			\bm y(t_0) & \bm y (t_{1}) & \dots & \bm y (t_{M}) \\
			\bm y (t_{-1}) & \bm y (t_{0}) & \dots & \bm y (t_{M-1})\\
			\vdots & \vdots &  &\vdots\\
			\bm y (t_{-N}) & \bm y (t_{1-N}) & \dots & \bm y (t_{M-N})
		\end{smallmatrix}
	\end{large}
	\right]
	= \left[
	\begin{large}
		\begin{smallmatrix}
			| && | \\[0.5ex]
			\bm \chi(t_0) & \dots & \bm \chi(t_M) \\[0.5ex]
			|  && |
		\end{smallmatrix}
	\end{large}
	\right]	 \,,
\end{equation}
where the rows $\bm \chi(t) \in \mathbb{R}^{(N+1)n_y}$ are the delay coordinate vectors at time $t$.
The number of delays $N$ is a hyperparameter, similar to an estimation horizon.
We assume observability with respect to the internal Koopman coordinates,
i.e., we can infer $\bm z$ from $\bm \chi$.
Next, we drop the full-state transformation $\bm \Psi(\bm x(t_0))$ in \cref{eqn:Koopman_observer} and formulate the modified variant:
\begin{subequations}
	\label{eqn:final}
	\begin{flalign}
		\dot{\bm z}(t) &= A \bm z(t) + B \bm u(t)  \,, \label{eqn:final_dyn} \\
		\left[\begin{large}\begin{smallmatrix}\bm x(t) \\ \bm y(t)
		\end{smallmatrix}\end{large}\right]
		&= \hat{\bm T}(\bm z(t))\,,
		\label{eqn:final_decode}
		\\
		\bm z(t_0) &= \hat{\bm \Psi}(\bm \chi(t_0)) \,. \label{eqn:estimator}
	\end{flalign}
\end{subequations}
Herein, $\hat{\bm \Psi} (\,\cdot\,)$ maps from the delay-coordinates $\bm \chi$ to the Koopman canonical coordinates $\bm z$ and obviates the need for a state observer. 
We refer to \cite{Casdagli.1991} for a further discussion of observability and delay embedding. 

The model (\ref{eqn:final}) is related to classical nonlinear autoregressive modeling for control, e.g., \cite{Draeger.1995}.
However, the main difference is that \cref{eqn:final} is strictly divided into the components linear dynamics and static nonlinear transformations, wherefore $\hat{\bm \Psi}$ can be excluded from an optimization problem.
Furthermore, autoregressive modeling strategies, including the Koopman delay embedding in
\cite{Korda.2018,Brunton.2017}, predict future $\bm \chi$ rather than the full  $(\bm x,\bm y)$.

Depending on the degree of nonlinearity of the system, the decoding in Eq.~(\ref{eqn:final_decode}) may be simplified using a matrix $C$ to obtain a linear approximator as prediction model:
\begin{subequations}
	\label{eqn:final_linear}
	\begin{flalign}
		\dot{\bm z}(t) &= A \bm z(t) + B \bm u(t)  \,,  \\
		\left[\begin{large}\begin{smallmatrix}\bm x(t) \\ \bm y(t)
		\end{smallmatrix}\end{large}\right]
		&= C \bm z(t)\,,\\
		\bm z(t_0) &= \hat{\bm \Psi}(\bm \chi(t_0)) \,. \label{eqn:estimator_linear}
	\end{flalign}
\end{subequations}

While \cref{eqn:estimator,eqn:estimator_linear} do not feature explicit parameters to incorporate statistical information about noise, 
the main advantage of this structure is the integration of model reduction and estimator design.
Both steps can be incorporated offline in a single identification pipeline, e.g., using deep learning.
Information on the reliability of combined estimation and prediction are extracted in the validation step. 

\section{Deep-learning Model Reduction Framework}
\label{sec:deeplearning}
To construct low-order models of the proposed form (\ref{eqn:final}), we adapt the deep-learning strategy
from our previous work \cite{Schulze.2022} with modifications regarding model structure, data set and training loss.
We train the reduced models on snapshots sampled from numerical simulations of the full-order model (e.g., a digital twin).
Due to the discrete nature of sampled data, 
training a discrete-time model form of Eq.~(\ref{eqn:final}) with zeroth-order hold is more straightforward:
\begin{subequations}
    \label{eqn:final_disc}
	\begin{flalign}
		\bm z_{k+1} &= \underline{A} \bm z_k + \underline{B} \bm u_k  \,, \\
		\left[\begin{large}\begin{smallmatrix}\bm x_{k} \\ \bm y_{k}
		\end{smallmatrix}\end{large}\right]
		&= \hat{\bm T}(\bm z_{k})\,,
		\\
		\bm z_0 &= \hat{\bm \Psi}(\bm \chi_0) \,.
	\end{flalign}
\end{subequations}
Notice that the underlying Koopman framework in Section~\ref{sec:review} is only valid for the continuous-time treatment, cf.~\cite{Schulze.2022}, but the linear dynamics of the Wiener form permit an exact transformation between discrete and continuous time.
Fig.~\ref{fig:deeplearning} depicts the Koopman network structure.
We use ANNs to learn suitable mappings
$\hat{\bm \Psi}$ (encoding) and $\hat{\bm T}$ (decoding).
The training loss $\mathcal{C}$ is computed as the sum of mean squared error terms for single and multi-time-step prediction: 
\begin{equation}
    \begin{split}
		\mathcal{C} \;=\;\; \frac{1}{s-1}& \sum_{k=0}^{s-1} \left\| 
		\begin{minimatrix}\bm x_{k+1}\\\bm y_{k+1}\end{minimatrix} 
		- \hat{\bm T}(\bm z_{k+1}(\bm \chi_{k}))  \right\|_{\mathrm{MSE}} \\
		 +\, \frac{1}{s-1}& \sum_{k=0}^{s-1} 
		\left\| 
		\begin{minimatrix}\bm x_{k+1}\\\bm y_{k+1}\end{minimatrix} 
		- \hat{\bm T}(\bm z_{k+1}(\bm \chi_{0})) \right\|_{\mathrm{MSE}}\\
		\mathrm{where:\phantom{xx}}
		\bm z_{k+1} &= \underline{A} \bm z_{k} + \underline{B} \bm u_{k} \,, \; k=j,j+1,...\\
		\bm z_j &= \bm \hat{\bm \Psi}(\bm \chi_j) \,.
    \end{split}
\end{equation}Therein, $s$ is the number of snapshots per trajectory, and
$\underline{A}$ and $\underline{B}$ are the discrete time system matrices.
The training problem can be easily adapted to models of the form (\ref{eqn:final_linear}).

Knowledge about the system may be used to preselect a (block) diagonal structure of $\underline{A}$, which promotes learning Koopman eigenfunctions and improves well-posedness of  
the training problem \cite{Schulze.2022}. 
While we do not consider measurement noise here, we could account for noise in the training, e.g., by adding Bayesian regularization. 
We implement the training framework using Python 3.9 and Tensorflow 2.5. 
\begin{figure}[h!]
	\centering
	\scalebox{0.85}{
		\begin{tikzpicture}[
			x=6ex, y=6ex, node distance=1ex and 5ex,
			squarednode/.style={rectangle, draw=black!100, fill=white!0, thin, minimum width=10ex, minimum height=6ex},]
			
			\node[draw, trapezium, 
			trapezium left angle = 70, trapezium right angle = 70,
			black, rotate=-90, trapezium stretches body, 
			text width=1cm, align=center] at (0,0) (encoder) {\rotatebox{90}{\parbox[c]{13ex}{\centering Encoder\\[1ex] \footnotesize$\bm z_0=\hat{\bm \Psi}(\, \cdot \,)$}}};
			
			\node[squarednode, align=center, minimum width=10ex] at (3,0)      (dynamic) {Linear\\dynamics};
			
			\node[draw, trapezium, 
			trapezium left angle = 70, trapezium right angle = 70,
			black, rotate=90, trapezium stretches body, 
			text width=1.cm, align=center] at (6,0) (decoder) {\rotatebox{-90}{\parbox[c]{13ex}{\centering Decoder\\[1ex] \footnotesize$
						\left[
						\begin{smallmatrix}
							\bm x_{k}\\\bm y_{k}
						\end{smallmatrix} \right]
						=\hat{\bm T}(\bm z_{k})$}}};
			
			\draw[-latex,] ($(encoder.south) + (-1.,1.)$) -- ($(encoder.south) + (0.,1.)$)
			node [pos=0.5,above,font=\small] {$\bm y_0$};
			\draw[-latex,] ($(encoder.south) + (-1.,.3)$) -- ($(encoder.south) + (0.,.3)$)
			node [pos=0.5,above,font=\small] {$\bm y_{-1}$};
			\draw[-latex,] ($(encoder.south) + (-1.,-0.3)$) -- ($(encoder.south) + (0.,-.3)$)
			node [pos=0.5,above,font=\small] {$\dots$};
			\draw[-latex,] ($(encoder.south) + (-1.,-1.)$) -- 
			($(encoder.south) + (0.,-1.)$)
			node [pos=0.5,above,font=\small] {$\bm y_{-N}$};
			
			\draw[-latex] (encoder.north) -- (dynamic.west)
			node [pos=0.5,above,font=\small] {$\bm z_0$};

			\draw[-latex] (dynamic.east) -- (decoder.north)
			node [pos=0.5,above,font=\small] {$\bm z_k$};
			\draw[-latex] (decoder.south) -- ($(decoder.south) + (1.2,0.)$) 
			node [pos=0.5,above,font=\small] {$\bm x_k ,\, \bm y_k$};
			
			\draw[-latex] ($(dynamic.south) - (0.,0.6)$) -- (dynamic.south) 
			node [pos=0.4,right,font=\small] {$\bm u_k$};
		\end{tikzpicture}
	}
	\caption{Network structure of the reduced models.}
	\label{fig:deeplearning}
\end{figure}
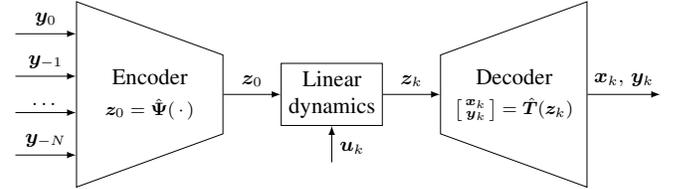

\begin{figure*}
	\centering
	\footnotesize
	\begin{minipage}{0.99\linewidth}
		\begin{minipage}[b]{0.325\linewidth}   
			\begin{tikzpicture}
				\begin{axis}[
				    compat=1.3,
					width = 1.0\linewidth,
					height = 0.8\linewidth,
					xmin=0, xmax=4.3,
					ymin=0.02, ymax=100,
					xmode=normal,
					ymode=log,
					scaled y ticks = false,
					log ticks with fixed point,
					xtick distance = 1,
					xlabel = {Time / h},
					ylabel = {Product impurity / ppm},
					legend cell align = {left},
					legend style={at={(1,1)},xshift=-2.cm,yshift=0.7cm, anchor=north west,nodes=right,legend columns=-1}
					]
					
					\addplot[blue, densely dashed, ultra thick] coordinates 
					{
						(0.000,6.176e+00)(0.033,6.176e+00)(0.067,6.176e+00)(0.100,6.176e+00)(0.133,6.176e+00)(0.167,6.176e+00)(0.200,6.176e+00)(0.233,6.176e+00)(0.267,6.176e+00)(0.300,6.176e+00)(0.333,6.176e+00)(0.367,1.035e+01)(0.400,1.469e+01)(0.433,2.056e+01)(0.467,2.640e+01)(0.500,3.085e+01)(0.533,3.396e+01)(0.567,3.609e+01)(0.600,3.754e+01)(0.633,3.850e+01)(0.667,3.906e+01)(0.700,3.931e+01)(0.733,3.930e+01)(0.767,3.907e+01)(0.800,3.866e+01)(0.833,3.810e+01)(0.867,3.741e+01)(0.900,3.662e+01)(0.933,3.575e+01)(0.967,3.481e+01)(1.000,3.383e+01)(1.033,3.282e+01)(1.067,3.178e+01)(1.100,3.074e+01)(1.133,2.970e+01)(1.167,2.866e+01)(1.200,2.764e+01)(1.233,2.664e+01)(1.267,2.566e+01)(1.300,2.471e+01)(1.333,2.378e+01)(1.367,2.289e+01)(1.400,2.204e+01)(1.433,2.121e+01)(1.467,2.042e+01)(1.500,1.966e+01)(1.533,1.894e+01)(1.567,1.825e+01)(1.600,1.759e+01)(1.633,1.697e+01)(1.667,1.637e+01)(1.700,1.581e+01)(1.733,1.527e+01)(1.767,1.476e+01)(1.800,1.428e+01)(1.833,1.382e+01)(1.867,1.339e+01)(1.900,1.298e+01)(1.933,1.259e+01)(1.967,1.222e+01)(2.000,1.188e+01)(2.033,1.155e+01)(2.067,1.124e+01)(2.100,1.095e+01)(2.133,1.068e+01)(2.167,1.042e+01)(2.200,1.017e+01)(2.233,9.939e+00)(2.267,9.720e+00)(2.300,9.515e+00)(2.333,9.320e+00)(2.367,7.456e+00)(2.400,6.199e+00)(2.433,5.197e+00)(2.467,4.365e+00)(2.500,3.659e+00)(2.533,3.054e+00)(2.567,2.537e+00)(2.600,2.100e+00)(2.633,1.734e+00)(2.667,1.432e+00)(2.700,1.185e+00)(2.733,9.867e-01)(2.767,8.283e-01)(2.800,7.031e-01)(2.833,6.050e-01)(2.867,5.285e-01)(2.900,4.693e-01)(2.933,4.236e-01)(2.967,3.886e-01)(3.000,3.618e-01)(3.033,3.413e-01)(3.067,3.257e-01)(3.100,3.139e-01)(3.133,3.049e-01)(3.167,2.981e-01)(3.200,2.929e-01)(3.233,2.890e-01)(3.267,2.860e-01)(3.300,2.838e-01)(3.333,2.821e-01)(3.367,2.808e-01)(3.400,2.798e-01)(3.433,2.791e-01)(3.467,2.785e-01)(3.500,2.781e-01)(3.533,2.778e-01)(3.567,2.776e-01)(3.600,2.774e-01)(3.633,2.772e-01)(3.667,2.771e-01)(3.700,2.771e-01)(3.733,2.770e-01)(3.767,2.770e-01)(3.800,2.769e-01)(3.833,2.769e-01)(3.867,2.769e-01)(3.900,2.769e-01)(3.933,2.769e-01)(3.967,2.768e-01)(4.000,2.768e-01)(4.033,2.768e-01)(4.067,2.768e-01)(4.100,2.768e-01)(4.133,2.768e-01)(4.167,2.768e-01)(4.200,2.768e-01)(4.233,2.768e-01)(4.267,2.768e-01)(4.300,2.768e-01)
					};

					\addplot[orange, densely dotted, ultra thick] coordinates 
					{						(0.000,5.955e+00)(0.033,5.955e+00)(0.067,5.888e+00)(0.100,5.904e+00)(0.133,5.901e+00)(0.167,5.891e+00)(0.200,5.885e+00)(0.233,5.884e+00)(0.267,5.889e+00)(0.300,5.901e+00)(0.333,5.919e+00)(0.367,1.002e+01)(0.400,1.921e+01)(0.433,2.406e+01)(0.467,2.583e+01)(0.500,2.618e+01)(0.533,2.596e+01)(0.567,2.549e+01)(0.600,2.494e+01)(0.633,2.434e+01)(0.667,2.374e+01)(0.700,2.314e+01)(0.733,2.254e+01)(0.767,2.196e+01)(0.800,2.140e+01)(0.833,2.086e+01)(0.867,2.035e+01)(0.900,1.985e+01)(0.933,1.938e+01)(0.967,1.893e+01)(1.000,1.850e+01)(1.033,1.809e+01)(1.067,1.770e+01)(1.100,1.734e+01)(1.133,1.699e+01)(1.167,1.666e+01)(1.200,1.635e+01)(1.233,1.606e+01)(1.267,1.578e+01)(1.300,1.551e+01)(1.333,1.526e+01)(1.367,1.503e+01)(1.400,1.480e+01)(1.433,1.459e+01)(1.467,1.439e+01)(1.500,1.420e+01)(1.533,1.401e+01)(1.567,1.384e+01)(1.600,1.368e+01)(1.633,1.352e+01)(1.667,1.337e+01)(1.700,1.323e+01)(1.733,1.310e+01)(1.767,1.297e+01)(1.800,1.285e+01)(1.833,1.273e+01)(1.867,1.262e+01)(1.900,1.251e+01)(1.933,1.241e+01)(1.967,1.231e+01)(2.000,1.222e+01)(2.033,1.213e+01)(2.067,1.204e+01)(2.100,1.196e+01)(2.133,1.188e+01)(2.167,1.180e+01)(2.200,1.173e+01)(2.233,1.166e+01)(2.267,1.159e+01)(2.300,1.152e+01)(2.333,1.146e+01)(2.367,8.533e+00)(2.400,7.105e+00)(2.433,5.758e+00)(2.467,4.634e+00)(2.500,3.742e+00)(2.533,3.043e+00)(2.567,2.497e+00)(2.600,2.066e+00)(2.633,1.725e+00)(2.667,1.452e+00)(2.700,1.232e+00)(2.733,1.053e+00)(2.767,9.068e-01)(2.800,7.864e-01)(2.833,6.866e-01)(2.867,6.033e-01)(2.900,5.333e-01)(2.933,4.742e-01)(2.967,4.239e-01)(3.000,3.810e-01)(3.033,3.441e-01)(3.067,3.123e-01)(3.100,2.847e-01)(3.133,2.606e-01)(3.167,2.396e-01)(3.200,2.211e-01)(3.233,2.048e-01)(3.267,1.903e-01)(3.300,1.775e-01)(3.333,1.660e-01)(3.367,1.558e-01)(3.400,1.466e-01)(3.433,1.383e-01)(3.467,1.309e-01)(3.500,1.241e-01)(3.533,1.180e-01)(3.567,1.124e-01)(3.600,1.073e-01)(3.633,1.027e-01)(3.667,9.842e-02)(3.700,9.451e-02)(3.733,9.092e-02)(3.767,8.761e-02)(3.800,8.455e-02)(3.833,8.172e-02)(3.867,7.910e-02)(3.900,7.667e-02)(3.933,7.441e-02)(3.967,7.231e-02)(4.000,7.035e-02)(4.033,6.853e-02)(4.067,6.682e-02)(4.100,6.522e-02)(4.133,6.372e-02)(4.167,6.232e-02)(4.200,6.101e-02)(4.233,5.977e-02)(4.267,5.860e-02)
					};
					
					\addplot[gray, dashdotted, ultra thick] coordinates 
					{						(0.000,6.168e+00)(0.033,6.168e+00)(0.067,6.285e+00)(0.100,6.267e+00)(0.133,6.259e+00)(0.167,6.261e+00)(0.200,6.272e+00)(0.233,6.289e+00)(0.267,6.312e+00)(0.300,6.339e+00)(0.333,6.370e+00)(0.367,1.091e+01)(0.400,1.756e+01)(0.433,2.354e+01)(0.467,2.772e+01)(0.500,3.003e+01)(0.533,3.092e+01)(0.567,3.089e+01)(0.600,3.031e+01)(0.633,2.946e+01)(0.667,2.849e+01)(0.700,2.749e+01)(0.733,2.651e+01)(0.767,2.558e+01)(0.800,2.470e+01)(0.833,2.388e+01)(0.867,2.312e+01)(0.900,2.242e+01)(0.933,2.177e+01)(0.967,2.117e+01)(1.000,2.062e+01)(1.033,2.011e+01)(1.067,1.963e+01)(1.100,1.919e+01)(1.133,1.878e+01)(1.167,1.840e+01)(1.200,1.804e+01)(1.233,1.771e+01)(1.267,1.740e+01)(1.300,1.711e+01)(1.333,1.684e+01)(1.367,1.659e+01)(1.400,1.635e+01)(1.433,1.613e+01)(1.467,1.592e+01)(1.500,1.572e+01)(1.533,1.554e+01)(1.567,1.536e+01)(1.600,1.520e+01)(1.633,1.505e+01)(1.667,1.490e+01)(1.700,1.477e+01)(1.733,1.464e+01)(1.767,1.452e+01)(1.800,1.440e+01)(1.833,1.430e+01)(1.867,1.420e+01)(1.900,1.410e+01)(1.933,1.401e+01)(1.967,1.392e+01)(2.000,1.384e+01)(2.033,1.377e+01)(2.067,1.370e+01)(2.100,1.363e+01)(2.133,1.357e+01)(2.167,1.351e+01)(2.200,1.345e+01)(2.233,1.340e+01)(2.267,1.335e+01)(2.300,1.330e+01)(2.333,1.326e+01)(2.367,9.688e+00)(2.400,8.376e+00)(2.433,6.982e+00)(2.467,5.716e+00)(2.500,4.652e+00)(2.533,3.790e+00)(2.567,3.103e+00)(2.600,2.559e+00)(2.633,2.128e+00)(2.667,1.785e+00)(2.700,1.510e+00)(2.733,1.289e+00)(2.767,1.110e+00)(2.800,9.626e-01)(2.833,8.415e-01)(2.867,7.409e-01)(2.900,6.567e-01)(2.933,5.859e-01)(2.967,5.258e-01)(3.000,4.746e-01)(3.033,4.307e-01)(3.067,3.929e-01)(3.100,3.600e-01)(3.133,3.315e-01)(3.167,3.064e-01)(3.200,2.844e-01)(3.233,2.650e-01)(3.267,2.478e-01)(3.300,2.325e-01)(3.333,2.188e-01)(3.367,2.066e-01)(3.400,1.956e-01)(3.433,1.857e-01)(3.467,1.767e-01)(3.500,1.686e-01)(3.533,1.612e-01)(3.567,1.545e-01)(3.600,1.483e-01)(3.633,1.427e-01)(3.667,1.375e-01)(3.700,1.328e-01)(3.733,1.284e-01)(3.767,1.244e-01)(3.800,1.207e-01)(3.833,1.172e-01)(3.867,1.140e-01)(3.900,1.110e-01)(3.933,1.083e-01)(3.967,1.057e-01)(4.000,1.033e-01)(4.033,1.011e-01)(4.067,9.898e-02)(4.100,9.702e-02)(4.133,9.519e-02)(4.167,9.347e-02)(4.200,9.185e-02)(4.233,9.033e-02)(4.267,8.891e-02)
					};
					
						\addplot[red, thick] coordinates 
					{						(0.000,6.155e+00)(0.033,6.155e+00)(0.067,6.137e+00)(0.100,6.157e+00)(0.133,6.176e+00)(0.167,6.192e+00)(0.200,6.205e+00)(0.233,6.214e+00)(0.267,6.220e+00)(0.300,6.225e+00)(0.333,6.229e+00)(0.367,8.493e+00)(0.400,1.660e+01)(0.433,2.169e+01)(0.467,2.743e+01)(0.500,3.240e+01)(0.533,3.634e+01)(0.567,3.918e+01)(0.600,4.104e+01)(0.633,4.212e+01)(0.667,4.261e+01)(0.700,4.263e+01)(0.733,4.232e+01)(0.767,4.173e+01)(0.800,4.095e+01)(0.833,4.001e+01)(0.867,3.895e+01)(0.900,3.779e+01)(0.933,3.655e+01)(0.967,3.527e+01)(1.000,3.395e+01)(1.033,3.261e+01)(1.067,3.128e+01)(1.100,2.996e+01)(1.133,2.868e+01)(1.167,2.744e+01)(1.200,2.624e+01)(1.233,2.511e+01)(1.267,2.404e+01)(1.300,2.302e+01)(1.333,2.207e+01)(1.367,2.119e+01)(1.400,2.036e+01)(1.433,1.959e+01)(1.467,1.888e+01)(1.500,1.821e+01)(1.533,1.760e+01)(1.567,1.704e+01)(1.600,1.651e+01)(1.633,1.603e+01)(1.667,1.558e+01)(1.700,1.516e+01)(1.733,1.478e+01)(1.767,1.443e+01)(1.800,1.410e+01)(1.833,1.380e+01)(1.867,1.352e+01)(1.900,1.326e+01)(1.933,1.302e+01)(1.967,1.280e+01)(2.000,1.259e+01)(2.033,1.240e+01)(2.067,1.222e+01)(2.100,1.206e+01)(2.133,1.191e+01)(2.167,1.177e+01)(2.200,1.164e+01)(2.233,1.152e+01)(2.267,1.140e+01)(2.300,1.130e+01)(2.333,1.120e+01)(2.367,7.662e+00)(2.400,6.664e+00)(2.433,5.780e+00)(2.467,4.920e+00)(2.500,4.142e+00)(2.533,3.470e+00)(2.567,2.904e+00)(2.600,2.433e+00)(2.633,2.043e+00)(2.667,1.723e+00)(2.700,1.460e+00)(2.733,1.245e+00)(2.767,1.069e+00)(2.800,9.260e-01)(2.833,8.089e-01)(2.867,7.130e-01)(2.900,6.343e-01)(2.933,5.695e-01)(2.967,5.160e-01)(3.000,4.716e-01)(3.033,4.347e-01)(3.067,4.040e-01)(3.100,3.783e-01)(3.133,3.566e-01)(3.167,3.384e-01)(3.200,3.231e-01)(3.233,3.101e-01)(3.267,2.991e-01)(3.300,2.897e-01)(3.333,2.818e-01)(3.367,2.750e-01)(3.400,2.693e-01)(3.433,2.644e-01)(3.467,2.602e-01)(3.500,2.567e-01)(3.533,2.537e-01)(3.567,2.512e-01)(3.600,2.491e-01)(3.633,2.474e-01)(3.667,2.460e-01)(3.700,2.449e-01)(3.733,2.440e-01)(3.767,2.433e-01)(3.800,2.428e-01)(3.833,2.424e-01)(3.867,2.423e-01)(3.900,2.422e-01)(3.933,2.423e-01)(3.967,2.425e-01)(4.000,2.427e-01)(4.033,2.431e-01)(4.067,2.435e-01)(4.100,2.440e-01)(4.133,2.446e-01)(4.167,2.452e-01)(4.200,2.458e-01)(4.233,2.465e-01)(4.267,2.472e-01)
					};

					\addlegendentry{Digital twin};
					\addlegendentry{Koopman Linear ($n_z$ = 10)};
					\addlegendentry{Koopman Linear ($n_z$ = 50)};
					\addlegendentry{Koopman Wiener ($n_z$ = 10)};
				\end{axis}
				\draw[] (0,0)  (-1.,-0.7) node[above,right] {\normalsize (a)};
			\end{tikzpicture}
		\end{minipage}	
		\hfill
		\begin{minipage}[b]{0.325\linewidth}  
			\begin{tikzpicture}
				\begin{axis}[
				    compat=1.3,
					width = 1.0\linewidth,
					height = 0.8\linewidth,
					xmin=0, xmax=4.3,
					ymin=150, ymax=210,
					xmode=normal,
					ymode=normal,
					xtick distance = 1,
					ytick distance = 25,
					xlabel = {Time / h},
					ylabel = {Production rate / mol$\,$s$^{-1}$},
					legend cell align = {left},
					legend style={at={(0.5,1)},xshift=0cm,yshift=0.6cm, anchor=north ,nodes=right,legend columns=-1}
					]
					
					\addplot[blue, densely dashed, ultra thick] coordinates 
					{						(0.000,155.166)(0.033,155.166)(0.067,155.166)(0.100,155.166)(0.133,155.166)(0.167,155.166)(0.200,155.166)(0.233,155.166)(0.267,155.166)(0.300,155.166)(0.333,155.166)(0.367,205.264)(0.400,207.071)(0.433,207.134)(0.467,207.131)(0.500,207.120)(0.533,207.108)(0.567,207.096)(0.600,207.083)(0.633,207.069)(0.667,207.055)(0.700,207.040)(0.733,207.026)(0.767,207.011)(0.800,206.997)(0.833,206.983)(0.867,206.971)(0.900,206.959)(0.933,206.948)(0.967,206.938)(1.000,206.929)(1.033,206.921)(1.067,206.914)(1.100,206.909)(1.133,206.904)(1.167,206.899)(1.200,206.896)(1.233,206.893)(1.267,206.891)(1.300,206.889)(1.333,206.887)(1.367,206.886)(1.400,206.885)(1.433,206.885)(1.467,206.884)(1.500,206.884)(1.533,206.884)(1.567,206.884)(1.600,206.884)(1.633,206.884)(1.667,206.884)(1.700,206.884)(1.733,206.885)(1.767,206.885)(1.800,206.885)(1.833,206.885)(1.867,206.886)(1.900,206.886)(1.933,206.886)(1.967,206.887)(2.000,206.887)(2.033,206.887)(2.067,206.887)(2.100,206.888)(2.133,206.888)(2.167,206.888)(2.200,206.888)(2.233,206.889)(2.267,206.889)(2.300,206.889)(2.333,206.889)(2.367,198.276)(2.400,198.511)(2.433,198.462)(2.467,198.413)(2.500,198.370)(2.533,198.342)(2.567,198.324)(2.600,198.315)(2.633,198.310)(2.667,198.308)(2.700,198.308)(2.733,198.309)(2.767,198.309)(2.800,198.310)(2.833,198.311)(2.867,198.312)(2.900,198.312)(2.933,198.312)(2.967,198.313)(3.000,198.313)(3.033,198.313)(3.067,198.313)(3.100,198.313)(3.133,198.313)(3.167,198.314)(3.200,198.314)(3.233,198.314)(3.267,198.314)(3.300,198.314)(3.333,198.314)(3.367,198.314)(3.400,198.314)(3.433,198.314)(3.467,198.314)(3.500,198.314)(3.533,198.314)(3.567,198.314)(3.600,198.314)(3.633,198.314)(3.667,198.314)(3.700,198.314)(3.733,198.314)(3.767,198.314)(3.800,198.314)(3.833,198.314)(3.867,198.314)(3.900,198.314)(3.933,198.314)(3.967,198.314)(4.000,198.314)(4.033,198.314)(4.067,198.314)(4.100,198.314)(4.133,198.314)(4.167,198.314)(4.200,198.314)(4.233,198.314)(4.267,198.314)(4.300,198.314)
					};
					
					\addplot[orange, densely dotted, ultra thick] coordinates 
					{						(0.000,155.221)(0.033,155.221)(0.067,154.932)(0.100,154.969)(0.133,154.989)(0.167,155.002)(0.200,155.013)(0.233,155.022)(0.267,155.030)(0.300,155.038)(0.333,155.046)(0.367,205.711)(0.400,206.268)(0.433,206.414)(0.467,206.457)(0.500,206.463)(0.533,206.455)(0.567,206.444)(0.600,206.433)(0.633,206.422)(0.667,206.413)(0.700,206.406)(0.733,206.399)(0.767,206.394)(0.800,206.389)(0.833,206.385)(0.867,206.382)(0.900,206.380)(0.933,206.377)(0.967,206.376)(1.000,206.374)(1.033,206.373)(1.067,206.373)(1.100,206.372)(1.133,206.372)(1.167,206.372)(1.200,206.371)(1.233,206.371)(1.267,206.371)(1.300,206.372)(1.333,206.372)(1.367,206.372)(1.400,206.372)(1.433,206.372)(1.467,206.373)(1.500,206.373)(1.533,206.373)(1.567,206.373)(1.600,206.374)(1.633,206.374)(1.667,206.374)(1.700,206.374)(1.733,206.375)(1.767,206.375)(1.800,206.375)(1.833,206.375)(1.867,206.375)(1.900,206.375)(1.933,206.376)(1.967,206.376)(2.000,206.376)(2.033,206.376)(2.067,206.376)(2.100,206.376)(2.133,206.376)(2.167,206.376)(2.200,206.376)(2.233,206.376)(2.267,206.376)(2.300,206.376)(2.333,206.376)(2.367,199.848)(2.400,199.820)(2.433,199.857)(2.467,199.873)(2.500,199.883)(2.533,199.890)(2.567,199.897)(2.600,199.903)(2.633,199.909)(2.667,199.915)(2.700,199.921)(2.733,199.927)(2.767,199.932)(2.800,199.937)(2.833,199.942)(2.867,199.947)(2.900,199.951)(2.933,199.955)(2.967,199.959)(3.000,199.963)(3.033,199.966)(3.067,199.969)(3.100,199.972)(3.133,199.974)(3.167,199.977)(3.200,199.979)(3.233,199.981)(3.267,199.983)(3.300,199.984)(3.333,199.986)(3.367,199.987)(3.400,199.988)(3.433,199.989)(3.467,199.990)(3.500,199.991)(3.533,199.991)(3.567,199.992)(3.600,199.992)(3.633,199.993)(3.667,199.993)(3.700,199.993)(3.733,199.993)(3.767,199.993)(3.800,199.993)(3.833,199.993)(3.867,199.993)(3.900,199.993)(3.933,199.993)(3.967,199.992)(4.000,199.992)(4.033,199.992)(4.067,199.992)(4.100,199.991)(4.133,199.991)(4.167,199.991)(4.200,199.990)(4.233,199.990)(4.267,199.990)
					};
					
					\addplot[gray, dashdotted, ultra thick] coordinates 
					{						(0.000,155.258)(0.033,155.258)(0.067,155.347)(0.100,155.346)(0.133,155.336)(0.167,155.327)(0.200,155.320)(0.233,155.315)(0.267,155.311)(0.300,155.307)(0.333,155.304)(0.367,205.627)(0.400,206.739)(0.433,206.819)(0.467,206.741)(0.500,206.665)(0.533,206.613)(0.567,206.580)(0.600,206.559)(0.633,206.546)(0.667,206.538)(0.700,206.533)(0.733,206.530)(0.767,206.529)(0.800,206.527)(0.833,206.526)(0.867,206.525)(0.900,206.524)(0.933,206.523)(0.967,206.522)(1.000,206.521)(1.033,206.519)(1.067,206.518)(1.100,206.517)(1.133,206.515)(1.167,206.513)(1.200,206.511)(1.233,206.509)(1.267,206.508)(1.300,206.505)(1.333,206.503)(1.367,206.501)(1.400,206.499)(1.433,206.497)(1.467,206.495)(1.500,206.492)(1.533,206.490)(1.567,206.487)(1.600,206.485)(1.633,206.483)(1.667,206.480)(1.700,206.478)(1.733,206.475)(1.767,206.473)(1.800,206.470)(1.833,206.468)(1.867,206.466)(1.900,206.463)(1.933,206.461)(1.967,206.458)(2.000,206.456)(2.033,206.453)(2.067,206.451)(2.100,206.448)(2.133,206.446)(2.167,206.444)(2.200,206.441)(2.233,206.439)(2.267,206.436)(2.300,206.434)(2.333,206.432)(2.367,200.207)(2.400,200.157)(2.433,200.125)(2.467,200.100)(2.500,200.083)(2.533,200.072)(2.567,200.066)(2.600,200.062)(2.633,200.060)(2.667,200.059)(2.700,200.059)(2.733,200.059)(2.767,200.059)(2.800,200.059)(2.833,200.059)(2.867,200.059)(2.900,200.059)(2.933,200.058)(2.967,200.058)(3.000,200.058)(3.033,200.057)(3.067,200.057)(3.100,200.056)(3.133,200.056)(3.167,200.055)(3.200,200.054)(3.233,200.054)(3.267,200.053)(3.300,200.052)(3.333,200.052)(3.367,200.051)(3.400,200.050)(3.433,200.049)(3.467,200.049)(3.500,200.048)(3.533,200.047)(3.567,200.046)(3.600,200.046)(3.633,200.045)(3.667,200.044)(3.700,200.044)(3.733,200.043)(3.767,200.042)(3.800,200.042)(3.833,200.041)(3.867,200.040)(3.900,200.040)(3.933,200.039)(3.967,200.039)(4.000,200.038)(4.033,200.037)(4.067,200.037)(4.100,200.036)(4.133,200.036)(4.167,200.035)(4.200,200.035)(4.233,200.034)(4.267,200.034)
					};
					
					\addplot[red, thick] coordinates 
					{						(0.000,155.193)(0.033,155.193)(0.067,155.178)(0.100,155.216)(0.133,155.241)(0.167,155.257)(0.200,155.267)(0.233,155.275)(0.267,155.280)(0.300,155.284)(0.333,155.287)(0.367,203.621)(0.400,207.394)(0.433,207.790)(0.467,207.724)(0.500,207.528)(0.533,207.279)(0.567,207.040)(0.600,206.835)(0.633,206.676)(0.667,206.563)(0.700,206.493)(0.733,206.458)(0.767,206.450)(0.800,206.462)(0.833,206.486)(0.867,206.517)(0.900,206.552)(0.933,206.588)(0.967,206.624)(1.000,206.659)(1.033,206.691)(1.067,206.721)(1.100,206.749)(1.133,206.775)(1.167,206.798)(1.200,206.820)(1.233,206.839)(1.267,206.857)(1.300,206.874)(1.333,206.889)(1.367,206.902)(1.400,206.915)(1.433,206.926)(1.467,206.936)(1.500,206.946)(1.533,206.954)(1.567,206.962)(1.600,206.969)(1.633,206.976)(1.667,206.982)(1.700,206.987)(1.733,206.992)(1.767,206.997)(1.800,207.001)(1.833,207.004)(1.867,207.008)(1.900,207.011)(1.933,207.013)(1.967,207.016)(2.000,207.018)(2.033,207.020)(2.067,207.021)(2.100,207.023)(2.133,207.024)(2.167,207.025)(2.200,207.026)(2.233,207.027)(2.267,207.028)(2.300,207.029)(2.333,207.029)(2.367,200.561)(2.400,199.835)(2.433,199.530)(2.467,199.388)(2.500,199.320)(2.533,199.276)(2.567,199.231)(2.600,199.175)(2.633,199.109)(2.667,199.036)(2.700,198.963)(2.733,198.893)(2.767,198.830)(2.800,198.775)(2.833,198.730)(2.867,198.694)(2.900,198.666)(2.933,198.647)(2.967,198.635)(3.000,198.628)(3.033,198.627)(3.067,198.631)(3.100,198.637)(3.133,198.647)(3.167,198.658)(3.200,198.672)(3.233,198.686)(3.267,198.702)(3.300,198.718)(3.333,198.735)(3.367,198.753)(3.400,198.770)(3.433,198.788)(3.467,198.805)(3.500,198.822)(3.533,198.839)(3.567,198.856)(3.600,198.872)(3.633,198.887)(3.667,198.902)(3.700,198.917)(3.733,198.930)(3.767,198.943)(3.800,198.955)(3.833,198.967)(3.867,198.977)(3.900,198.987)(3.933,198.996)(3.967,199.004)(4.000,199.011)(4.033,199.018)(4.067,199.023)(4.100,199.028)(4.133,199.032)(4.167,199.035)(4.200,199.037)(4.233,199.039)(4.267,199.039)
					};

				\end{axis}
				\draw[] (0,0)  (-1.,-0.7) node[above,right] {\normalsize (b)};
			\end{tikzpicture}
		\end{minipage}
		\vspace{2ex}
    \hfill
		\begin{minipage}[b]{0.325\linewidth}  
			\begin{tikzpicture}
				\begin{axis}[
				    compat=1.3,
					width = 1.0\linewidth,
					height = 0.8\linewidth,
					xmin=0, xmax=4.3,
					ymin=0.0002, ymax=0.2,
					xmode=normal,
					ymode=log,
					scaled y ticks = false,
					log ticks with fixed point,
					xtick distance = 1,
					xlabel = {Time / h},
					ylabel = {Mole fraction $1-x_{N_2,20}$ / $-$},
					legend cell align = {left},
					legend pos = north east,
					]
					
					\addplot[blue, densely dashed, ultra thick] coordinates 
					{
						(0.000,1.221e-02)(0.033,1.221e-02)(0.067,1.221e-02)(0.100,1.221e-02)(0.133,1.221e-02)(0.167,1.221e-02)(0.200,1.221e-02)(0.233,1.221e-02)(0.267,1.221e-02)(0.300,1.221e-02)(0.333,1.221e-02)(0.367,7.270e-02)(0.400,1.079e-01)(0.433,1.078e-01)(0.467,1.024e-01)(0.500,9.676e-02)(0.533,9.162e-02)(0.567,8.703e-02)(0.600,8.293e-02)(0.633,7.924e-02)(0.667,7.588e-02)(0.700,7.280e-02)(0.733,6.995e-02)(0.767,6.729e-02)(0.800,6.478e-02)(0.833,6.242e-02)(0.867,6.017e-02)(0.900,5.802e-02)(0.933,5.597e-02)(0.967,5.399e-02)(1.000,5.209e-02)(1.033,5.026e-02)(1.067,4.850e-02)(1.100,4.680e-02)(1.133,4.517e-02)(1.167,4.359e-02)(1.200,4.208e-02)(1.233,4.062e-02)(1.267,3.922e-02)(1.300,3.788e-02)(1.333,3.659e-02)(1.367,3.535e-02)(1.400,3.417e-02)(1.433,3.304e-02)(1.467,3.196e-02)(1.500,3.093e-02)(1.533,2.994e-02)(1.567,2.900e-02)(1.600,2.811e-02)(1.633,2.725e-02)(1.667,2.644e-02)(1.700,2.567e-02)(1.733,2.494e-02)(1.767,2.424e-02)(1.800,2.358e-02)(1.833,2.295e-02)(1.867,2.235e-02)(1.900,2.179e-02)(1.933,2.125e-02)(1.967,2.075e-02)(2.000,2.027e-02)(2.033,1.981e-02)(2.067,1.938e-02)(2.100,1.898e-02)(2.133,1.859e-02)(2.167,1.823e-02)(2.200,1.788e-02)(2.233,1.756e-02)(2.267,1.725e-02)(2.300,1.696e-02)(2.333,1.669e-02)(2.367,1.623e-02)(2.400,1.380e-02)(2.433,1.141e-02)(2.467,9.324e-03)(2.500,7.607e-03)(2.533,6.224e-03)(2.567,5.121e-03)(2.600,4.249e-03)(2.633,3.564e-03)(2.667,3.029e-03)(2.700,2.615e-03)(2.733,2.295e-03)(2.767,2.050e-03)(2.800,1.862e-03)(2.833,1.718e-03)(2.867,1.609e-03)(2.900,1.526e-03)(2.933,1.463e-03)(2.967,1.415e-03)(3.000,1.379e-03)(3.033,1.351e-03)(3.067,1.331e-03)(3.100,1.315e-03)(3.133,1.303e-03)(3.167,1.294e-03)(3.200,1.287e-03)(3.233,1.282e-03)(3.267,1.278e-03)(3.300,1.275e-03)(3.333,1.273e-03)(3.367,1.271e-03)(3.400,1.270e-03)(3.433,1.269e-03)(3.467,1.268e-03)(3.500,1.268e-03)(3.533,1.268e-03)(3.567,1.267e-03)(3.600,1.267e-03)(3.633,1.267e-03)(3.667,1.267e-03)(3.700,1.267e-03)(3.733,1.267e-03)(3.767,1.266e-03)(3.800,1.266e-03)(3.833,1.266e-03)(3.867,1.266e-03)(3.900,1.266e-03)(3.933,1.266e-03)(3.967,1.266e-03)(4.000,1.266e-03)(4.033,1.266e-03)(4.067,1.266e-03)(4.100,1.266e-03)(4.133,1.266e-03)(4.167,1.266e-03)(4.200,1.266e-03)(4.233,1.266e-03)(4.267,1.266e-03)(4.300,1.266e-03)
					};
					
					\addplot[orange, densely dotted, ultra thick] coordinates 
					{						(0.000,1.243e-02)(0.033,1.243e-02)(0.067,1.251e-02)(0.100,1.259e-02)(0.133,1.266e-02)(0.167,1.272e-02)(0.200,1.278e-02)(0.233,1.284e-02)(0.267,1.290e-02)(0.300,1.296e-02)(0.333,1.302e-02)(0.367,5.858e-02)(0.400,7.533e-02)(0.433,7.362e-02)(0.467,6.784e-02)(0.500,6.191e-02)(0.533,5.666e-02)(0.567,5.217e-02)(0.600,4.836e-02)(0.633,4.512e-02)(0.667,4.234e-02)(0.700,3.995e-02)(0.733,3.788e-02)(0.767,3.608e-02)(0.800,3.450e-02)(0.833,3.310e-02)(0.867,3.187e-02)(0.900,3.077e-02)(0.933,2.979e-02)(0.967,2.891e-02)(1.000,2.812e-02)(1.033,2.741e-02)(1.067,2.676e-02)(1.100,2.617e-02)(1.133,2.563e-02)(1.167,2.514e-02)(1.200,2.469e-02)(1.233,2.428e-02)(1.267,2.389e-02)(1.300,2.354e-02)(1.333,2.321e-02)(1.367,2.291e-02)(1.400,2.262e-02)(1.433,2.236e-02)(1.467,2.211e-02)(1.500,2.188e-02)(1.533,2.167e-02)(1.567,2.146e-02)(1.600,2.127e-02)(1.633,2.109e-02)(1.667,2.092e-02)(1.700,2.076e-02)(1.733,2.061e-02)(1.767,2.046e-02)(1.800,2.032e-02)(1.833,2.019e-02)(1.867,2.007e-02)(1.900,1.995e-02)(1.933,1.984e-02)(1.967,1.973e-02)(2.000,1.963e-02)(2.033,1.953e-02)(2.067,1.943e-02)(2.100,1.934e-02)(2.133,1.926e-02)(2.167,1.917e-02)(2.200,1.909e-02)(2.233,1.901e-02)(2.267,1.894e-02)(2.300,1.887e-02)(2.333,1.880e-02)(2.367,1.656e-02)(2.400,1.349e-02)(2.433,1.102e-02)(2.467,9.083e-03)(2.500,7.573e-03)(2.533,6.386e-03)(2.567,5.444e-03)(2.600,4.689e-03)(2.633,4.077e-03)(2.667,3.576e-03)(2.700,3.162e-03)(2.733,2.818e-03)(2.767,2.528e-03)(2.800,2.283e-03)(2.833,2.074e-03)(2.867,1.895e-03)(2.900,1.740e-03)(2.933,1.606e-03)(2.967,1.489e-03)(3.000,1.386e-03)(3.033,1.296e-03)(3.067,1.215e-03)(3.100,1.144e-03)(3.133,1.080e-03)(3.167,1.023e-03)(3.200,9.722e-04)(3.233,9.260e-04)(3.267,8.841e-04)(3.300,8.462e-04)(3.333,8.116e-04)(3.367,7.801e-04)(3.400,7.513e-04)(3.433,7.249e-04)(3.467,7.006e-04)(3.500,6.783e-04)(3.533,6.577e-04)(3.567,6.386e-04)(3.600,6.210e-04)(3.633,6.046e-04)(3.667,5.894e-04)(3.700,5.752e-04)(3.733,5.621e-04)(3.767,5.497e-04)(3.800,5.382e-04)(3.833,5.275e-04)(3.867,5.174e-04)(3.900,5.079e-04)(3.933,4.991e-04)(3.967,4.907e-04)(4.000,4.829e-04)(4.033,4.755e-04)(4.067,4.685e-04)(4.100,4.619e-04)(4.133,4.557e-04)(4.167,4.498e-04)(4.200,4.443e-04)(4.233,4.390e-04)(4.267,4.340e-04)
					};
					
					\addplot[gray, dashdotted, ultra thick] coordinates 
					{						(0.000,1.238e-02)(0.033,1.238e-02)(0.067,1.304e-02)(0.100,1.297e-02)(0.133,1.294e-02)(0.167,1.294e-02)(0.200,1.296e-02)(0.233,1.298e-02)(0.267,1.302e-02)(0.300,1.307e-02)(0.333,1.312e-02)(0.367,5.797e-02)(0.400,8.338e-02)(0.433,7.719e-02)(0.467,6.752e-02)(0.500,5.992e-02)(0.533,5.433e-02)(0.567,5.010e-02)(0.600,4.680e-02)(0.633,4.412e-02)(0.667,4.188e-02)(0.700,3.998e-02)(0.733,3.833e-02)(0.767,3.688e-02)(0.800,3.560e-02)(0.833,3.447e-02)(0.867,3.345e-02)(0.900,3.253e-02)(0.933,3.171e-02)(0.967,3.096e-02)(1.000,3.028e-02)(1.033,2.967e-02)(1.067,2.911e-02)(1.100,2.859e-02)(1.133,2.813e-02)(1.167,2.770e-02)(1.200,2.730e-02)(1.233,2.694e-02)(1.267,2.661e-02)(1.300,2.630e-02)(1.333,2.602e-02)(1.367,2.576e-02)(1.400,2.551e-02)(1.433,2.529e-02)(1.467,2.508e-02)(1.500,2.488e-02)(1.533,2.470e-02)(1.567,2.453e-02)(1.600,2.437e-02)(1.633,2.422e-02)(1.667,2.408e-02)(1.700,2.395e-02)(1.733,2.383e-02)(1.767,2.371e-02)(1.800,2.360e-02)(1.833,2.349e-02)(1.867,2.339e-02)(1.900,2.329e-02)(1.933,2.320e-02)(1.967,2.311e-02)(2.000,2.303e-02)(2.033,2.295e-02)(2.067,2.287e-02)(2.100,2.279e-02)(2.133,2.272e-02)(2.167,2.265e-02)(2.200,2.258e-02)(2.233,2.252e-02)(2.267,2.245e-02)(2.300,2.239e-02)(2.333,2.232e-02)(2.367,1.909e-02)(2.400,1.629e-02)(2.433,1.336e-02)(2.467,1.102e-02)(2.500,9.226e-03)(2.533,7.831e-03)(2.567,6.727e-03)(2.600,5.841e-03)(2.633,5.118e-03)(2.667,4.522e-03)(2.700,4.026e-03)(2.733,3.610e-03)(2.767,3.257e-03)(2.800,2.956e-03)(2.833,2.698e-03)(2.867,2.476e-03)(2.900,2.282e-03)(2.933,2.114e-03)(2.967,1.966e-03)(3.000,1.836e-03)(3.033,1.721e-03)(3.067,1.618e-03)(3.100,1.527e-03)(3.133,1.445e-03)(3.167,1.372e-03)(3.200,1.306e-03)(3.233,1.246e-03)(3.267,1.192e-03)(3.300,1.143e-03)(3.333,1.098e-03)(3.367,1.057e-03)(3.400,1.019e-03)(3.433,9.846e-04)(3.467,9.528e-04)(3.500,9.235e-04)(3.533,8.964e-04)(3.567,8.713e-04)(3.600,8.481e-04)(3.633,8.264e-04)(3.667,8.063e-04)(3.700,7.875e-04)(3.733,7.700e-04)(3.767,7.536e-04)(3.800,7.382e-04)(3.833,7.238e-04)(3.867,7.102e-04)(3.900,6.975e-04)(3.933,6.855e-04)(3.967,6.742e-04)(4.000,6.635e-04)(4.033,6.534e-04)(4.067,6.439e-04)(4.100,6.348e-04)(4.133,6.263e-04)(4.167,6.181e-04)(4.200,6.104e-04)(4.233,6.030e-04)(4.267,5.961e-04)
					};

					\addplot[red, thick] coordinates 
					{
						(0.000,1.226e-02)(0.033,1.226e-02)(0.067,1.221e-02)(0.100,1.221e-02)(0.133,1.222e-02)(0.167,1.224e-02)(0.200,1.226e-02)(0.233,1.228e-02)(0.267,1.230e-02)(0.300,1.232e-02)(0.333,1.234e-02)(0.367,8.716e-02)(0.400,1.200e-01)(0.433,1.141e-01)(0.467,1.078e-01)(0.500,1.024e-01)(0.533,9.809e-02)(0.567,9.382e-02)(0.600,8.961e-02)(0.633,8.557e-02)(0.667,8.179e-02)(0.700,7.826e-02)(0.733,7.497e-02)(0.767,7.190e-02)(0.800,6.899e-02)(0.833,6.620e-02)(0.867,6.352e-02)(0.900,6.091e-02)(0.933,5.838e-02)(0.967,5.590e-02)(1.000,5.350e-02)(1.033,5.117e-02)(1.067,4.893e-02)(1.100,4.678e-02)(1.133,4.473e-02)(1.167,4.280e-02)(1.200,4.097e-02)(1.233,3.926e-02)(1.267,3.767e-02)(1.300,3.618e-02)(1.333,3.480e-02)(1.367,3.352e-02)(1.400,3.233e-02)(1.433,3.124e-02)(1.467,3.023e-02)(1.500,2.930e-02)(1.533,2.845e-02)(1.567,2.766e-02)(1.600,2.693e-02)(1.633,2.626e-02)(1.667,2.564e-02)(1.700,2.507e-02)(1.733,2.455e-02)(1.767,2.406e-02)(1.800,2.362e-02)(1.833,2.320e-02)(1.867,2.282e-02)(1.900,2.247e-02)(1.933,2.214e-02)(1.967,2.184e-02)(2.000,2.156e-02)(2.033,2.131e-02)(2.067,2.107e-02)(2.100,2.085e-02)(2.133,2.064e-02)(2.167,2.045e-02)(2.200,2.028e-02)(2.233,2.012e-02)(2.267,1.996e-02)(2.300,1.982e-02)(2.333,1.970e-02)(2.367,1.687e-02)(2.400,1.482e-02)(2.433,1.260e-02)(2.467,1.059e-02)(2.500,8.865e-03)(2.533,7.421e-03)(2.567,6.226e-03)(2.600,5.249e-03)(2.633,4.458e-03)(2.667,3.821e-03)(2.700,3.312e-03)(2.733,2.906e-03)(2.767,2.581e-03)(2.800,2.321e-03)(2.833,2.112e-03)(2.867,1.944e-03)(2.900,1.808e-03)(2.933,1.698e-03)(2.967,1.608e-03)(3.000,1.535e-03)(3.033,1.475e-03)(3.067,1.427e-03)(3.100,1.386e-03)(3.133,1.354e-03)(3.167,1.327e-03)(3.200,1.305e-03)(3.233,1.287e-03)(3.267,1.273e-03)(3.300,1.262e-03)(3.333,1.253e-03)(3.367,1.246e-03)(3.400,1.240e-03)(3.433,1.236e-03)(3.467,1.234e-03)(3.500,1.232e-03)(3.533,1.231e-03)(3.567,1.231e-03)(3.600,1.231e-03)(3.633,1.232e-03)(3.667,1.234e-03)(3.700,1.235e-03)(3.733,1.237e-03)(3.767,1.239e-03)(3.800,1.242e-03)(3.833,1.244e-03)(3.867,1.247e-03)(3.900,1.250e-03)(3.933,1.252e-03)(3.967,1.255e-03)(4.000,1.258e-03)(4.033,1.261e-03)(4.067,1.263e-03)(4.100,1.266e-03)(4.133,1.269e-03)(4.167,1.271e-03)(4.200,1.274e-03)(4.233,1.276e-03)(4.267,1.279e-03)
					};
				\end{axis}
				\draw[] (0,0)  (-1.,-0.7) node[above,right] {\normalsize (c)};
			\end{tikzpicture}
		\end{minipage}
	\end{minipage}	
	\caption{Open-loop model step test: (a) product impurity, (b) production rate, (c) residual molar fraction on tray 20.}
	\label{fig:test}
\end{figure*}
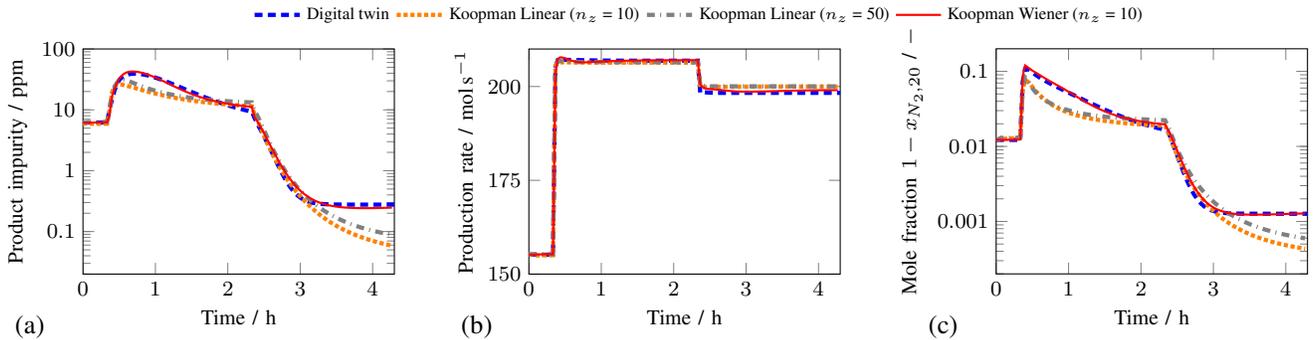
\vspace{-2ex}
\section{NMPC Case Study}
\label{sec:casestudy}
Distillation columns are among the core units of many chemical plants, such as cryogenic air separation units \cite{Pattison.2017}.
For plant operation,
NMPC is considered as a promising strategy to address the inherent nonlinearities and strong variable coupling. 
However, the high computational effort for optimizing detailed column models and other units frequently prohibits real-time NMPC \cite{CaspariWave, Schulze.2021}. 
Here, we employ our model reduction strategy to enable real-time NMPC of the cryogenic distillation column discussed in \cite{CaspariWave, Pattison.2017}.

The column has 45 equilibrium separation trays and fractionates air to produce gaseous nitrogen (N$_2$) withdrawn at the top.
The feed air stream enters the column bottom at \unit[100]{K} and \unit[5.5]{bar} with vapor feed rate $F$.
Liquid reflux $R$ at the column top is provided by means of a total condenser.
The available measurements for state estimation are the molar flows of vapor product $V$ and bottom liquid stream $L$, and the temperature $T_{10}$ on the 10th tray from the bottom.
The control degrees of freedom are the reflux ratio, $\xi=R/(R+V)$, and feed rate $F$.
The controlled variables are the product impurity, $p = 1 - x_{N_2,prod}$, 
and product flow rate $V$.
Notice that the product impurity is controlled but not measured.
This assumption is fairly ambitious, but viable due to the high accuracy of the proposed control models.

We start from a high-fidelity digital twin of the column, i.e., a physical model similar to \cite{Pattison.2017}. 
In particular, we model air as ternary nitrogen-argon-oxygen mixture, formulate tray-by-tray material and energy balances, linear hydraulic correlations, and use Margules activity model and extended Antoine equations for thermodynamic computations.
The full differential-algebraic model has $135$ differential and $2140$ algebraic equations.
The model is a nonlinear semi-explicit differential-algebraic equation system (DAE) of index one.
The system is asymptotically stable, and we assume that 
the DAE has a unique and smooth solution and
resembles the behavior of \cref{eqn:inputaffine}.
All simulations and optimizations are performed with our open-source dynamic optimization software DyOS \cite{CaspariDyOS} using SNOPT \cite{SNOPT}.
We specify tolerances of $10^{-8}$ for all simulations, and $10^{-5}$ for feasibility and optimality.
All computations run on a desktop computer with Intel Core i5-8500 CPU at 3.0 GHz and 16 GB RAM.

\subsection{Data sampling and model identification}
To generate the trajectory snapshots, we simulated the digital twin subject to a sequence of 400 input steps.
Each input step has \unit[3]{h} duration and was randomly drawn from a set of uniformly distributed
control tuples of $\xi \in [0.51,0.54]$ and $V \in [200,400]\,\unit{mol/s}$.
Preliminary experimentation showed that a balanced amount of dynamic trajectories and stationary data is the most crucial ingredient to a successful model training.
In particular, too little stationary data caused stability issues and steady-state offset, whereas to much stationary data resulted in poor reproduction of the dynamic response.
Consequently, we enriched the dynamic step response data set explicitly with steady-state data as described below.

The sampling time is \unit[2]{min} and reflects the fastest relevant response of the stiff dynamics.
We specified $N$=$\,$20 for delay embedding. 
For the training, we log-transformed all molar fractions and scaled all variables between zero and one.
We built the training trajectory set by sliding along the recorded data 
in a moving horizon fashion, stopping every 10 sampling instants and copying 
$s\,$=$\,$60 consecutive snapshots (\unit[2]{h} training trajectories). Further, for each input combination, we added a pure steady-state trajectory of \unit[2]{h} length.
Finally, mini-batches of 32 trajectories were divided the into \unit[80]{\%} training and \unit[20]{\%} validation data.

A systematic parameter study suggested that dynamics of $n_z\,$=$\,$10 (versus 135 original states) are satisfactory, when paired with encoder and decoder with tanh activation and two hidden layers of (50, 20) and (20, 50) neurons, respectively.
Inspection of the sampled data showed an over-damped response and suggested to preset a diagonal structure of the matrix $\underline{A}$,
which proved to be applicable without a negative influence on the prediction quality in all trainings.

To benchmark the accuracy, we also trained linear models, \cref{eqn:final_linear}, by using the deep-learning framework, but removing the nonlinear hidden decoder layers.
All models were trained for $10\,000$ epochs using the optimizer Adam.
After training, we retrieved 
the weights with smallest validation loss.

\begin{figure*}
	\centering
	\scriptsize 
	\begin{minipage}{0.95\linewidth} 
		\begin{minipage}[b]{0.45\linewidth}  
			\begin{tikzpicture}
			\begin{axis}[
				width = 1.0\linewidth,
				height = 0.6\linewidth,
				xmin=0, xmax=4,
				ymin=90, ymax=210,
				xmode=normal,
				ymode=normal,
				xtick distance = 1,
				xlabel = {Time / h},
				ylabel = {Production rate / mol s$^{-1}$},
				legend cell align = {left},
				legend style={at={(0,1)},xshift=-1.7cm,yshift=0.7cm, anchor=north west,nodes=right,legend columns=-1}
				]

			\addplot[black, dotted, thick] coordinates {(0,200)(5,200)};
			
			\addplot[black, dashed,] coordinates {(0,150)(0.5,150)(0.5,187.5)(2.5,187.5)(2.5,112.5)(3.5,112.5)(3.5,150)(4,150)};

			\addplot[blue, dash pattern=on 8pt off 2pt, ultra thick] coordinates 
			{
				(0.033,150.000)(0.067,150.000)(0.100,150.000)(0.133,150.000)(0.167,150.000)(0.200,150.000)(0.233,150.000)(0.267,150.000)(0.300,150.000)(0.333,150.000)(0.367,150.000)(0.400,150.000)(0.433,150.000)(0.467,150.000)(0.500,150.000)(0.533,187.500)(0.567,187.500)(0.600,187.500)(0.633,187.500)(0.667,187.500)(0.700,187.500)(0.733,187.500)(0.767,187.500)(0.800,187.500)(0.833,187.500)(0.867,187.500)(0.900,187.500)(0.933,187.500)(0.967,187.500)(1.000,187.500)(1.033,187.500)(1.067,187.491)(1.100,187.500)(1.133,187.494)(1.167,187.500)(1.200,187.500)(1.233,187.500)(1.267,187.500)(1.300,187.500)(1.333,187.500)(1.367,187.500)(1.400,187.500)(1.433,187.500)(1.467,187.500)(1.500,187.500)(1.533,187.500)(1.567,187.500)(1.600,187.500)(1.633,187.500)(1.667,187.500)(1.700,187.500)(1.733,187.500)(1.767,187.500)(1.800,187.500)(1.833,187.500)(1.867,187.500)(1.900,187.500)(1.933,187.500)(1.967,187.500)(2.000,187.500)(2.033,187.500)(2.067,187.500)(2.100,187.500)(2.133,187.500)(2.167,187.500)(2.200,187.500)(2.233,187.500)(2.267,187.500)(2.300,187.500)(2.333,187.500)(2.367,187.500)(2.400,187.500)(2.433,187.500)(2.467,187.500)(2.500,187.500)(2.533,112.500)(2.567,112.500)(2.600,112.500)(2.633,112.500)(2.667,112.500)(2.700,112.500)(2.733,112.500)(2.767,112.500)(2.800,112.500)(2.833,112.500)(2.867,112.500)(2.900,112.500)(2.933,112.500)(2.967,112.500)(3.000,112.500)(3.033,112.500)(3.067,112.500)(3.100,112.500)(3.133,112.500)(3.167,112.500)(3.200,112.500)(3.233,112.500)(3.267,112.500)(3.300,112.500)(3.333,112.500)(3.367,112.500)(3.400,112.500)(3.433,112.500)(3.467,112.500)(3.500,112.500)(3.533,150.000)(3.567,150.000)(3.600,150.000)(3.633,150.000)(3.667,150.000)(3.700,150.000)(3.733,150.000)(3.767,150.000)(3.800,150.000)(3.833,150.000)(3.867,150.000)(3.900,150.000)(3.933,150.000)(3.967,150.000)(4.000,150.000)(4.033,150.000)
			};
			
			\addplot[red, very thick] coordinates 
			{
				(0.033,150.183)(0.067,150.307)(0.100,150.343)(0.133,150.343)(0.167,150.329)(0.200,150.313)(0.233,150.304)(0.267,150.294)(0.300,150.291)(0.333,150.289)(0.367,150.287)(0.400,150.286)(0.433,150.283)(0.467,150.280)(0.500,150.275)(0.533,188.007)(0.567,187.226)(0.600,187.072)(0.633,187.237)(0.667,187.201)(0.700,187.125)(0.733,187.079)(0.767,187.096)(0.800,187.082)(0.833,187.066)(0.867,187.181)(0.900,187.154)(0.933,187.150)(0.967,187.147)(1.000,187.203)(1.033,187.254)(1.067,187.319)(1.100,187.360)(1.133,187.376)(1.167,187.396)(1.200,187.448)(1.233,187.666)(1.267,187.600)(1.300,187.537)(1.333,187.514)(1.367,187.500)(1.400,187.488)(1.433,187.477)(1.467,187.469)(1.500,187.464)(1.533,187.463)(1.567,187.474)(1.600,187.519)(1.633,187.554)(1.667,187.585)(1.700,187.613)(1.733,187.644)(1.767,187.675)(1.800,187.701)(1.833,187.732)(1.867,187.771)(1.900,187.812)(1.933,187.850)(1.967,187.885)(2.000,187.918)(2.033,187.956)(2.067,187.979)(2.100,187.937)(2.133,187.937)(2.167,187.954)(2.200,187.983)(2.233,188.015)(2.267,188.047)(2.300,188.072)(2.333,188.088)(2.367,188.096)(2.400,188.086)(2.433,188.052)(2.467,187.990)(2.500,187.891)(2.533,113.209)(2.567,111.330)(2.600,111.489)(2.633,112.386)(2.667,112.684)(2.700,112.496)(2.733,112.213)(2.767,112.174)(2.800,112.163)(2.833,112.131)(2.867,112.165)(2.900,112.296)(2.933,112.355)(2.967,112.317)(3.000,112.489)(3.033,112.578)(3.067,112.685)(3.100,112.878)(3.133,113.015)(3.167,113.125)(3.200,113.102)(3.233,112.784)(3.267,112.767)(3.300,112.660)(3.333,112.631)(3.367,112.615)(3.400,112.604)(3.433,112.597)(3.467,112.592)(3.500,112.596)(3.533,149.768)(3.567,149.789)(3.600,150.321)(3.633,150.913)(3.667,151.087)(3.700,151.060)(3.733,151.007)(3.767,150.861)(3.800,150.651)(3.833,150.537)(3.867,150.521)(3.900,150.447)(3.933,150.393)(3.967,150.268)(4.000,150.188)(4.033,150.100)
			};
			
			\addplot[green, dash pattern=on 8pt off 2pt, thick] coordinates 
			{
				(0.033,149.850)(0.067,149.936)(0.100,149.908)(0.133,149.846)(0.167,149.792)(0.200,149.746)(0.233,149.707)(0.267,149.675)(0.300,149.650)(0.333,149.628)(0.367,149.610)(0.400,149.608)(0.433,149.613)(0.467,149.617)(0.500,149.625)(0.533,186.101)(0.567,187.250)(0.600,187.169)(0.633,187.252)(0.667,187.310)(0.700,187.264)(0.733,187.189)(0.767,187.217)(0.800,187.267)(0.833,187.067)(0.867,187.140)(0.900,187.182)(0.933,187.133)(0.967,187.112)(1.000,187.118)(1.033,187.099)(1.067,187.083)(1.100,187.168)(1.133,187.274)(1.167,187.312)(1.200,187.381)(1.233,187.193)(1.267,187.150)(1.300,187.192)(1.333,187.248)(1.367,187.303)(1.400,187.357)(1.433,187.408)(1.467,187.457)(1.500,187.504)(1.533,187.697)(1.567,187.776)(1.600,187.844)(1.633,187.911)(1.667,187.972)(1.700,188.026)(1.733,188.079)(1.767,188.125)(1.800,188.165)(1.833,188.198)(1.867,188.226)(1.900,188.246)(1.933,188.262)(1.967,188.285)(2.000,188.313)(2.033,188.353)(2.067,188.420)(2.100,188.514)(2.133,188.587)(2.167,188.528)(2.200,188.457)(2.233,188.483)(2.267,188.554)(2.300,188.597)(2.333,188.573)(2.367,188.499)(2.400,188.408)(2.433,188.324)(2.467,188.257)(2.500,188.203)(2.533,112.017)(2.567,112.395)(2.600,113.528)(2.633,112.233)(2.667,113.070)(2.700,113.392)(2.733,113.078)(2.767,112.591)(2.800,112.375)(2.833,112.186)(2.867,111.611)(2.900,112.067)(2.933,112.404)(2.967,112.502)(3.000,112.537)(3.033,112.597)(3.067,112.562)(3.100,112.536)(3.133,112.504)(3.167,112.518)(3.200,112.619)(3.233,112.468)(3.267,112.175)(3.300,112.077)(3.333,112.111)(3.367,112.161)(3.400,112.204)(3.433,112.228)(3.467,112.239)(3.500,112.248)(3.533,149.177)(3.567,149.877)(3.600,150.037)(3.633,150.347)(3.667,150.483)(3.700,150.539)(3.733,150.422)(3.767,150.162)(3.800,150.044)(3.833,149.523)(3.867,149.177)(3.900,149.304)(3.933,149.486)(3.967,149.633)(4.000,149.698)(4.033,149.747)
			};
					
					\addplot[gray, dash dot, very thick] coordinates{(0,0)(0,1)};
					\addplot[orange, dash dot, very thick] coordinates{(0,0)(0,1)};
					
					\addplot[black, dotted, thick] coordinates {(0,100)(5,100)};
					
					\addlegendentry{Constraints};
					\addlegendentry{Setpoint};
					\addlegendentry{Ideal NMPC};
					\addlegendentry{Koopman NMPC (Plant)};
					\addlegendentry{Koopman LMPC (Plant)};
        			\addlegendentry{Koopman NMPC (Prediction)};
        			\addlegendentry{Koopman LMPC (Prediction)};
					
				\end{axis}
				\draw[] (0,0) (-1.5,-0.1) node[above,right] {\normalsize (a)};
			\end{tikzpicture}
			
		\end{minipage}
	\hfill
		\begin{minipage}[b]{0.45\linewidth}   
			\begin{tikzpicture}
			\begin{axis}[
			width = 1.0\linewidth,
			height = 0.6\linewidth,
			xmin=0, xmax=4,
			xmode=normal,
			ymin=0.3, ymax=100,
			ymode=log,
			xtick distance = 1,
			scaled y ticks = false,
			log ticks with fixed point,
			xlabel = {Time / h},
			ylabel = {Impurity / ppm},
			legend cell align = {left},
			legend pos = south west,
			]

			\addplot[blue, dash pattern=on 8pt off 2pt, thick] coordinates 
			{
				(0.033,38.595)(0.067,37.418)(0.100,36.332)(0.133,35.228)(0.167,34.055)(0.200,32.814)(0.233,31.522)(0.267,30.210)(0.300,28.897)(0.333,27.592)(0.367,26.304)(0.400,25.041)(0.433,23.809)(0.467,22.618)(0.500,21.481)(0.533,30.147)(0.567,36.050)(0.600,42.336)(0.633,47.103)(0.667,49.490)(0.700,50.000)(0.733,49.443)(0.767,48.239)(0.800,46.630)(0.833,44.772)(0.867,42.770)(0.900,40.691)(0.933,38.583)(0.967,36.479)(1.000,34.403)(1.033,32.375)(1.067,30.405)(1.100,28.511)(1.133,26.693)(1.167,24.961)(1.200,23.317)(1.233,21.764)(1.267,20.300)(1.300,18.927)(1.333,17.642)(1.367,16.443)(1.400,15.327)(1.433,14.292)(1.467,13.333)(1.500,12.446)(1.533,11.628)(1.567,10.876)(1.600,10.187)(1.633,9.557)(1.667,8.985)(1.700,8.489)(1.733,8.076)(1.767,7.737)(1.800,7.454)(1.833,7.219)(1.867,7.030)(1.900,6.888)(1.933,6.792)(1.967,6.743)(2.000,6.742)(2.033,6.788)(2.067,6.881)(2.100,7.022)(2.133,7.210)(2.167,7.446)(2.200,7.734)(2.233,8.075)(2.267,8.473)(2.300,8.931)(2.333,9.454)(2.367,10.045)(2.400,10.710)(2.433,11.457)(2.467,12.297)(2.500,13.361)(2.533,6.538)(2.567,4.379)(2.600,3.069)(2.633,2.201)(2.667,1.599)(2.700,1.186)(2.733,0.915)(2.767,0.750)(2.800,0.660)(2.833,0.619)(2.867,0.608)(2.900,0.615)(2.933,0.631)(2.967,0.652)(3.000,0.674)(3.033,0.695)(3.067,0.714)(3.100,0.732)(3.133,0.747)(3.167,0.760)(3.200,0.771)(3.233,0.780)(3.267,0.788)(3.300,0.795)(3.333,0.801)(3.367,0.805)(3.400,0.810)(3.433,0.813)(3.467,0.816)(3.500,0.818)(3.533,1.462)(3.567,2.158)(3.600,3.220)(3.633,4.449)(3.667,5.436)(3.700,6.053)(3.733,6.345)(3.767,6.394)(3.800,6.269)(3.833,6.024)(3.867,5.701)(3.900,5.330)(3.933,4.938)(3.967,4.541)(4.000,4.152)(4.033,3.783)
			};
					
				\addplot[red, ultra thick] coordinates{
				(0.033,38.913)(0.067,37.873)(0.100,36.885)(0.133,35.876)(0.167,34.772)(0.200,33.543)(0.233,32.207)(0.267,30.796)(0.300,29.352)(0.333,27.907)(0.367,26.481)(0.400,25.091)(0.433,23.749)(0.467,22.471)(0.500,21.279)(0.533,30.029)(0.567,35.187)(0.600,40.715)(0.633,44.982)(0.667,46.842)(0.700,46.823)(0.733,45.737)(0.767,44.103)(0.800,42.129)(0.833,39.960)(0.867,37.738)(0.900,35.466)(0.933,33.214)(0.967,31.010)(1.000,28.891)(1.033,26.859)(1.067,24.929)(1.100,23.099)(1.133,21.369)(1.167,19.745)(1.200,18.231)(1.233,16.845)(1.267,15.534)(1.300,14.316)(1.333,13.192)(1.367,12.157)(1.400,11.207)(1.433,10.337)(1.467,9.544)(1.500,8.821)(1.533,8.165)(1.567,7.576)(1.600,7.068)(1.633,6.628)(1.667,6.252)(1.700,5.932)(1.733,5.666)(1.767,5.455)(1.800,5.296)(1.833,5.191)(1.867,5.147)(1.900,5.169)(1.933,5.263)(1.967,5.434)(2.000,5.694)(2.033,6.056)(2.067,6.525)(2.100,7.061)(2.133,7.685)(2.167,8.417)(2.200,9.274)(2.233,10.268)(2.267,11.404)(2.300,12.683)(2.333,14.106)(2.367,15.671)(2.400,17.362)(2.433,19.131)(2.467,20.900)(2.500,22.532)(2.533,11.755)(2.567,8.167)(2.600,5.980)(2.633,4.501)(2.667,3.416)(2.700,2.627)(2.733,2.089)(2.767,1.752)(2.800,1.561)(2.833,1.469)(2.867,1.444)(2.900,1.465)(2.933,1.513)(2.967,1.576)(3.000,1.650)(3.033,1.728)(3.067,1.806)(3.100,1.885)(3.133,1.959)(3.167,2.030)(3.200,2.091)(3.233,2.135)(3.267,2.175)(3.300,2.203)(3.333,2.223)(3.367,2.237)(3.400,2.246)(3.433,2.252)(3.467,2.256)(3.500,2.258)(3.533,3.924)(3.567,5.643)(3.600,8.209)(3.633,11.131)(3.667,13.567)(3.700,15.309)(3.733,16.453)(3.767,17.115)(3.800,17.397)(3.833,17.409)(3.867,17.230)(3.900,16.880)(3.933,16.413)(3.967,15.845)(4.000,15.218)(4.033,14.547)
			};
			
			\addplot[gray, dash dot, very thick] coordinates{
				(0.033,36.373)(0.067,40.063)(0.100,39.559)(0.133,38.438)(0.167,36.984)(0.200,35.404)(0.233,33.606)(0.267,31.811)(0.300,30.066)(0.333,28.413)(0.367,26.825)(0.400,25.291)(0.433,23.875)(0.467,22.557)(0.500,21.402)(0.533,30.016)(0.567,33.210)(0.600,44.692)(0.633,49.735)(0.667,49.994)(0.700,49.992)(0.733,49.999)(0.767,48.413)(0.800,46.776)(0.833,43.486)(0.867,39.287)(0.900,38.104)(0.933,35.860)(0.967,33.127)(1.000,30.641)(1.033,28.231)(1.067,26.202)(1.100,23.907)(1.133,21.512)(1.167,19.772)(1.200,18.687)(1.233,16.100)(1.267,14.743)(1.300,13.746)(1.333,12.772)(1.367,11.853)(1.400,11.005)(1.433,10.234)(1.467,9.527)(1.500,8.872)(1.533,8.256)(1.567,7.688)(1.600,7.213)(1.633,6.768)(1.667,6.370)(1.700,6.024)(1.733,5.733)(1.767,5.498)(1.800,5.314)(1.833,5.194)(1.867,5.145)(1.900,5.165)(1.933,5.253)(1.967,5.418)(2.000,5.663)(2.033,6.004)(2.067,6.417)(2.100,6.796)(2.133,7.370)(2.167,8.119)(2.200,8.984)(2.233,9.936)(2.267,10.954)(2.300,12.038)(2.333,13.218)(2.367,14.542)(2.400,16.013)(2.433,17.579)(2.467,19.131)(2.500,20.479)(2.533,11.260)(2.567,7.173)(2.600,5.705)(2.633,4.285)(2.667,3.338)(2.700,2.772)(2.733,2.156)(2.767,1.913)(2.800,1.752)(2.833,1.696)(2.867,1.573)(2.900,1.559)(2.933,1.528)(2.967,1.500)(3.000,1.562)(3.033,1.654)(3.067,1.718)(3.100,1.775)(3.133,1.863)(3.167,1.875)(3.200,1.914)(3.233,2.241)(3.267,2.192)(3.300,2.176)(3.333,2.200)(3.367,2.224)(3.400,2.238)(3.433,2.244)(3.467,2.248)(3.500,2.248)(3.533,3.490)(3.567,5.115)(3.600,8.456)(3.633,10.989)(3.667,13.031)(3.700,14.494)(3.733,15.381)(3.767,16.009)(3.800,16.213)(3.833,16.399)(3.867,16.344)(3.900,15.922)(3.933,15.459)(3.967,14.859)(4.000,14.396)(4.033,13.950)
			};
			
				\addplot[green, dash pattern=on 8pt off 2pt, ultra thick] coordinates{
				(0.033,39.200)(0.067,38.459)(0.100,37.715)(0.133,36.931)(0.167,36.065)(0.200,35.104)(0.233,34.045)(0.267,32.899)(0.300,31.671)(0.333,30.362)(0.367,28.968)(0.400,27.486)(0.433,25.901)(0.467,24.186)(0.500,22.291)(0.533,29.641)(0.567,34.119)(0.600,38.247)(0.633,40.591)(0.667,40.744)(0.700,39.383)(0.733,37.216)(0.767,34.724)(0.800,32.124)(0.833,29.472)(0.867,26.945)(0.900,24.539)(0.933,22.259)(0.967,20.133)(1.000,18.166)(1.033,16.352)(1.067,14.691)(1.100,13.189)(1.133,11.834)(1.167,10.609)(1.200,9.511)(1.233,8.512)(1.267,7.625)(1.300,6.841)(1.333,6.148)(1.367,5.538)(1.400,5.002)(1.433,4.532)(1.467,4.122)(1.500,3.764)(1.533,3.456)(1.567,3.197)(1.600,2.981)(1.633,2.804)(1.667,2.659)(1.700,2.542)(1.733,2.453)(1.767,2.389)(1.800,2.350)(1.833,2.335)(1.867,2.343)(1.900,2.376)(1.933,2.436)(1.967,2.528)(2.000,2.656)(2.033,2.830)(2.067,3.060)(2.100,3.356)(2.133,3.733)(2.167,4.179)(2.200,4.689)(2.233,5.280)(2.267,5.970)(2.300,6.773)(2.333,7.696)(2.367,8.739)(2.400,9.902)(2.433,11.180)(2.467,12.573)(2.500,14.075)(2.533,7.513)(2.567,5.471)(2.600,4.181)(2.633,3.183)(2.667,2.480)(2.700,1.976)(2.733,1.643)(2.767,1.460)(2.800,1.395)(2.833,1.411)(2.867,1.468)(2.900,1.569)(2.933,1.692)(2.967,1.826)(3.000,1.967)(3.033,2.113)(3.067,2.256)(3.100,2.394)(3.133,2.522)(3.167,2.642)(3.200,2.755)(3.233,2.850)(3.267,2.926)(3.300,2.994)(3.333,3.057)(3.367,3.116)(3.400,3.169)(3.433,3.219)(3.467,3.267)(3.500,3.313)(3.533,5.795)(3.567,8.474)(3.600,12.345)(3.633,16.579)(3.667,20.080)(3.700,22.677)(3.733,24.499)(3.767,25.700)(3.800,26.467)(3.833,26.756)(3.867,26.743)(3.900,26.610)(3.933,26.361)(3.967,26.018)(4.000,25.572)(4.033,25.055)
			};
			
			    \addplot[orange, dash dot, very thick] coordinates{
				(0.033,36.059)(0.067,39.072)(0.100,38.401)(0.133,37.123)(0.167,35.756)(0.200,34.379)(0.233,33.165)(0.267,31.953)(0.300,30.732)(0.333,29.469)(0.367,28.153)(0.400,26.769)(0.433,25.313)(0.467,23.742)(0.500,22.026)(0.533,30.096)(0.567,32.949)(0.600,36.899)(0.633,36.298)(0.667,37.888)(0.700,38.353)(0.733,37.387)(0.767,35.108)(0.800,31.005)(0.833,25.497)(0.867,24.449)(0.900,20.841)(0.933,19.465)(0.967,18.134)(1.000,17.270)(1.033,16.649)(1.067,15.990)(1.100,13.642)(1.133,11.498)(1.167,9.693)(1.200,8.423)(1.233,7.247)(1.267,7.228)(1.300,6.856)(1.333,6.102)(1.367,5.380)(1.400,4.746)(1.433,4.208)(1.467,3.762)(1.500,3.397)(1.533,3.098)(1.567,2.853)(1.600,2.650)(1.633,2.488)(1.667,2.357)(1.700,2.252)(1.733,2.172)(1.767,2.115)(1.800,2.081)(1.833,2.072)(1.867,2.086)(1.900,2.127)(1.933,2.192)(1.967,2.283)(2.000,2.406)(2.033,2.566)(2.067,2.768)(2.100,3.015)(2.133,3.347)(2.167,3.751)(2.200,4.188)(2.233,4.670)(2.267,5.204)(2.300,5.828)(2.333,6.623)(2.367,7.601)(2.400,8.699)(2.433,9.854)(2.467,11.048)(2.500,12.295)(2.533,5.641)(2.567,4.010)(2.600,3.606)(2.633,3.595)(2.667,2.318)(2.700,1.739)(2.733,1.546)(2.767,1.355)(2.800,1.335)(2.833,1.340)(2.867,1.656)(2.900,1.629)(2.933,1.656)(2.967,1.766)(3.000,1.906)(3.033,2.017)(3.067,2.134)(3.100,2.210)(3.133,2.303)(3.167,2.359)(3.200,2.725)(3.233,2.760)(3.267,2.800)(3.300,2.840)(3.333,2.858)(3.367,2.917)(3.400,2.991)(3.433,3.054)(3.467,3.109)(3.500,3.165)(3.533,4.809)(3.567,8.784)(3.600,12.152)(3.633,15.310)(3.667,18.799)(3.700,20.794)(3.733,22.172)(3.767,22.548)(3.800,23.357)(3.833,23.891)(3.867,23.487)(3.900,22.584)(3.933,22.508)(3.967,22.525)(4.000,22.895)(4.033,22.578)
			};
			
			\addplot[blue, dash pattern=on 8pt off 2pt,  thick] coordinates 
			{
				(0.033,38.595)(0.067,37.418)(0.100,36.332)(0.133,35.228)(0.167,34.055)(0.200,32.814)(0.233,31.522)(0.267,30.210)(0.300,28.897)(0.333,27.592)(0.367,26.304)(0.400,25.041)(0.433,23.809)(0.467,22.618)(0.500,21.481)(0.533,30.147)(0.567,36.050)(0.600,42.336)(0.633,47.103)(0.667,49.490)(0.700,50.000)(0.733,49.443)(0.767,48.239)(0.800,46.630)(0.833,44.772)(0.867,42.770)(0.900,40.691)(0.933,38.583)(0.967,36.479)(1.000,34.403)(1.033,32.375)(1.067,30.405)(1.100,28.511)(1.133,26.693)(1.167,24.961)(1.200,23.317)(1.233,21.764)(1.267,20.300)(1.300,18.927)(1.333,17.642)(1.367,16.443)(1.400,15.327)(1.433,14.292)(1.467,13.333)(1.500,12.446)(1.533,11.628)(1.567,10.876)(1.600,10.187)(1.633,9.557)(1.667,8.985)(1.700,8.489)(1.733,8.076)(1.767,7.737)(1.800,7.454)(1.833,7.219)(1.867,7.030)(1.900,6.888)(1.933,6.792)(1.967,6.743)(2.000,6.742)(2.033,6.788)(2.067,6.881)(2.100,7.022)(2.133,7.210)(2.167,7.446)(2.200,7.734)(2.233,8.075)(2.267,8.473)(2.300,8.931)(2.333,9.454)(2.367,10.045)(2.400,10.710)(2.433,11.457)(2.467,12.297)(2.500,13.361)(2.533,6.538)(2.567,4.379)(2.600,3.069)(2.633,2.201)(2.667,1.599)(2.700,1.186)(2.733,0.915)(2.767,0.750)(2.800,0.660)(2.833,0.619)(2.867,0.608)(2.900,0.615)(2.933,0.631)(2.967,0.652)(3.000,0.674)(3.033,0.695)(3.067,0.714)(3.100,0.732)(3.133,0.747)(3.167,0.760)(3.200,0.771)(3.233,0.780)(3.267,0.788)(3.300,0.795)(3.333,0.801)(3.367,0.805)(3.400,0.810)(3.433,0.813)(3.467,0.816)(3.500,0.818)(3.533,1.462)(3.567,2.158)(3.600,3.220)(3.633,4.449)(3.667,5.436)(3.700,6.053)(3.733,6.345)(3.767,6.394)(3.800,6.269)(3.833,6.024)(3.867,5.701)(3.900,5.330)(3.933,4.938)(3.967,4.541)(4.000,4.152)(4.033,3.783)
			};
			
			\addplot[black, dotted, thick] coordinates {(0,0.5)(5,0.5)};
			\addplot[black, dotted, thick] coordinates {(0,50)(5,50)};
			
			\end{axis}
			\draw[] (0,0)  (-1.1,-0.1) node[above,right] {\normalsize (b)};
			\end{tikzpicture}
		\end{minipage}

	\end{minipage}
	
	\caption{Closed-loop response of the controlled plant: (a) tracking of production rate, (b) satisfaction of purity constraints. }
	\label{fig:closedloop}
\end{figure*}
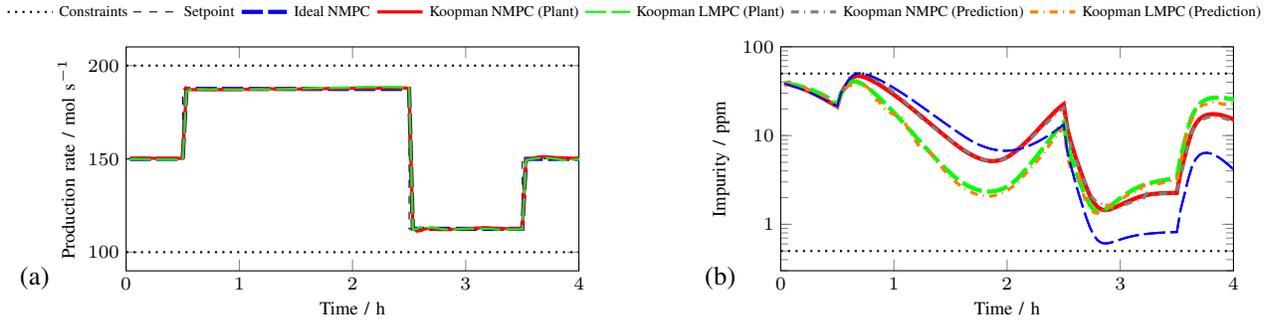
\subsection{Open-loop model test}
We evaluate the model in an independent test scenario.
The process is initialized at the stationary point corresponding to $(\xi,F)\,$=$\,(0.52,300)$, and subject to two consecutive \unit[1.5]{h} input steps to the input limits $(0.52,400)$ and $(0.54,400)$.
For model assessment, we perform a single simulation sweep given only $\bm\chi(t_0)$ and sampled $\bm u(t)$.

Fig.~\ref{fig:test}a compares the product impurity.
Despite the considerable excitation, the Wiener-type model predicts the system response precisely over several orders of magnitude at all times.
The linear Koopman model also captures the main trends.
However, its prediction exhibits a notable deviations from the original trajectory. The underestimation of purity grades may be critical if tight control of different quality grades is desired.
To assess the effect of a higher-dimensional lifting on the linear Koopman predictions, we trained more complex linear models using $n_z\,$=$\,$50, and (100, 75) hidden neurons for the nonlinear encoding.
The extended lifting improves the prediction slightly, but the prediction offset is still present, Fig.~\ref{fig:test}a.
Notice that while higher lifting reduces the prediction error, the state estimation problem becomes increasingly more complex.

Fig.~\ref{fig:test}b shows the response of the production rate and Fig.~\ref{fig:test}c exemplarily depicts the mole fraction of residual components, $1-x_{N_2}$, on tray 20. 
Again, the predictions by the Wiener-type Koopman model are precise with vanishing 
steady-state offset, whereas deviations in the linear prediction are more pronounced.

\subsection{Koopman LMPC and NMPC formulation}
To demonstrate that data-driven nonlinear model reduction using the proposed Koopman structure, Eq.~(\ref{eqn:final}), enables real-time NMPC,
we consider a controller solving:
\begin{subequations}
	\label{eqn:nmpc}
	\begin{flalign}
		\min_{\bm u} \sum_{k=1}^{N_c} &
		\ell_k(\bm x_k, \bm y_k) 
		\label{eqn:nmpccost}\\ 
		\mathrm{s.t.}\hspace{5ex}
		\bm z_{k+1} &= \underline{A} \bm z_k + \underline{B} \bm u_k  \,,
		\label{eqn:nmpc_koopman1}\\
		\left[\begin{large}\begin{smallmatrix}\bm x_{k+1} \\ \bm y_{k+1}
		\end{smallmatrix}\end{large}\right]
		&= \hat{\bm T}(\bm z_{k+1}) \,,
		\label{eqn:nmpc_koopman2}\\
		\bm z_0 &= \hat{\bm \Psi}(\bm \chi_0) \,, \label{eqn:initial}\\
		\bm x_{k+1} &\in \mathcal{X} ,\, \bm y_{k+1} \in \mathcal{Y} ,\; \bm u_{k} \in \mathcal{U}
		\,,\\
		k&=0,1,... , N_c - 1
		\,.
	\end{flalign}
\end{subequations}
Herein, $N_c$ is the control horizon, 
\cref{eqn:nmpccost} describes the 
cost function with stage cost $\ell_k$,
\cref{eqn:nmpc_koopman1,eqn:nmpc_koopman2} are the Koopman model,
and $\mathcal{X}$, $\mathcal{Y}$, $\mathcal{U}$ are the constrained admissible sets of states, outputs and controls, respectively.
We will refer to this controller as ``Koopman NMPC''.
The encoding of the plant measurements,
Eq.~(\ref{eqn:initial}), provides feedback and is evaluated prior to solving the optimization problem.
In addition, we investigate ``Koopman LMPC'' obtained by substituting \cref{eqn:nmpc_koopman1,eqn:nmpc_koopman2,eqn:initial} by the proposed linear reduced model, Eq.~(\ref{eqn:final_linear}), for $n_z\,$=$\,10$.

Here, we implement tracking controllers, 
where $\ell_k (\,\cdot\,) := (V_k - V_{sp,k})^2 $ penalizes the setpoint tracking error of the production rate.
The control moves are piecewise constant (zeroth-order hold), and consecutive optimizations are warmstart initialized.
We employ the digital twin as the controlled plant in our in-silico study.
To benchmark the closed-loop NMPC performance and CPU costs,
we setup an ``ideal NMPC'' optimizing the full-order digital twin in combination with full-state feedback.
We neglect the time delay introduced by solving the optimization problem. 
Table \ref{tab:tuning} summarizes the controller tuning parameters and constraints.
\begin{table}[ht]
	\centering
	\caption{NMPC parameters and constraints.} 
	\label{tab:tuning}
	\begin{small}
		\begin{tabular}{lcll}
			\toprule
			\textbf{Variable} & \textbf{Symbol} & \textbf{Value} & \textbf{Unit}\\
			\midrule
			Sampling time & $\Delta t_s$ & 2 &\unit{min}\\
			Control horizon & $t_f$ & 60 & \unit{min}\\
			Product impurity  & $p$& $[0.5,\,50]$ & $ \unit{ppm}$\\
			Production rate  & $V$& $[100,\,200]$ & $ \unit{mol\,s^{-1}}$\\
			Reflux ratio & $u_1 = \xi$& $[0.51,\,0.54]$ & $-$\\
			Feed rate & $u_2 = F$& $[200,\,400]$ & $\unit{mol\,s^{-1}}$\\
			\bottomrule
			\vspace{-5ex}
		\end{tabular}
	\end{small}
\end{table}
\subsection{Closed-loop results}
We show the results of the control study.
Initially, the plant is at steady state, where the impurity fraction is \unit[40]{ppm} and the production rate is $\unit[150]{mol\,s^{-1}}$.
The control task is to track a series of anticipated setpoint changes, Fig.~\ref{fig:closedloop}a, while maintaining feasible operation within the bounds, Fig.~\ref{fig:closedloop}b.

Figs.~\ref{fig:closedloop}a and \ref{fig:closedloop}b depict the closed-loop controlled plant response.
Further, \ref{fig:closedloop}b shows the series of MPC-internal first-step open-loop predictions to assess the interplay of state initialization and prediction versus the actual plant response.
All controllers successfully accomplish to track the product demand while satisfying the quality restrictions at all times.
The adjustments of the setpoint are tracked precisely and with no overshoot or steady-state offset, Fig.~\ref{fig:closedloop}a.
Minor deviations from the setpoint profile are only temporary and within an acceptable range (below $\unit[1]{\%}$). 
Importantly, all controllers do not violate the constraints.
Despite the slightly less accurate open-loop predictions of the linear models, see also prediction vs.~plant in \cref{fig:closedloop}b, process operation by Koopman LMPC is satisfactory.
While all controllers succeed to operate the process,
their product purity trajectories, Fig.~\ref{fig:closedloop}b, differ.
However, these differences are not associated with a loss, since the feasibility target is accomplished.

We next compare the performance in terms of the CPU costs for solving the optimization to assess the real-time capability,
\cref{tab:cpu}.
Although all setups provide similar tracking performance, 
the CPU times differ clearly.
Average and maximum CPU time of the benchmark NMPC are
two orders of magnitude greater than for Koopman NMPC and LMPC.
The resulting control delay of the benchmark NMPC lies above the sampling time and poses a serious issue in terms of closed-loop performance and stability. 
Conversely, the computational effort using reduced Koopman models is almost negligible with all CPU times lying in a narrow range below \unit[10]{sec}.
The corresponding average CPU time reduction is
\unit[99]{\%}.
While the CPU expenses for LMPC and NMPC are comparable, we expect a further LMPC speed-up when directly using a quadratic programming solver.
Further, we expect a speed-up of Koopman NMPC 
from exploitation of the Wiener block-structure in the sensitivity computation.

\begin{table}[h!]
	\centering
	\caption{Comparison of CPU times for solving the MPC optimization programs to convergence.}
	\label{tab:cpu}
	\begin{small}
		\begin{tabular}{lccc}
			\toprule
			\textbf{Controller} & \textbf{\footnotesize $\varnothing$ CPU time} & \textbf{\footnotesize Max.~CPU time} & \textbf{\footnotesize $\varnothing$ Red.} \\ 
			\midrule
			Benchmark NMPC & \unit[252]{sec} & \unit[409]{sec} & \\
			Koopman NMPC & \unit[3.4]{sec} & \unit[8]{sec} & \unit[99]{\%}\\ 
			Koopman LMPC & \unit[2.9]{sec} & \unit[4]{sec} & \unit[99]{\%}\\
			\bottomrule
		\end{tabular}
	\end{small}
\end{table}
\section{Conclusions and Outlook}
\label{sec:conclusion}
We present a data-driven model reduction strategy that generates linear as well as Wiener-type reduced models and incorporates state estimation features for control.
The approach is based on Koopman theory and equipped with delay-coordinate embedding 
for model initialization.
We employ a deep-learning framework for the simultaneous training of the reduced dynamics and the state estimation features. 

In a case study, we demonstrate that the proposed reduction approach enables both real-time NMPC and LMPC.
Therein, we consider production rate tracking of a nonlinear high-purity cryogenic distillation column.
We show that the proposed models with very low-dimensional latent linear dynamics succeed to reproduce the high-dimensional nonlinear dynamics. 
The models provide reliable initialization of the current states as well as accurate predictions.
This enables feasible operation and high tracking performance at \unit[99]{\%} average CPU time reduction compared to a controller using the full-order model.
While Koopman LMPC and NMPC offer similar controller performance here, the method of choice will generally depend on the control task and degree of nonlinearity of the controlled system.

We conclude that data-driven model reduction based on Koopman theory enables real-time NMPC and LMPC of a chemical process of industrial relevance. 
Future work will investigate online model improvement in closed-loop operation \cite{Hewing.2020}. 
Further, we will apply the framework to control more complex chemical plants, including air separation units.

\section*{ACKNOWLEDGMENT}
The authors gratefully acknowledge the  financial support of the Kopernikus project SynErgie by the Federal Ministry of Education and Research (BMBF) and project supervision by the project management organization Projekttr\"ager J\"ulich.

\bibliographystyle{ieeetr}

\end{document}